\RequirePackage{fix-cm}
\documentclass[smallcondensed]{svjour3}     
\smartqed  
\usepackage{graphicx}

\usepackage{cite}
\usepackage{hyperref}
\usepackage{amsmath,amssymb,amsfonts}

\usepackage{algorithmic}
\usepackage[ruled,vlined]{algorithm2e}
\usepackage{graphicx}
\usepackage{subcaption}
\usepackage{multirow}
\usepackage{textcomp}
\usepackage{xcolor}
\usepackage{framed}
\usepackage{amsthm}

\newtheorem{thm}{Theorem}
\newcommand{\commentout}[1]{}

\newcommand{\real}{\mathbb{R}}

\newcommand{\pre}{pre}
\newcommand{\con}{con}

\newcommand{\pcriterion}{\textbf{P-rule}}
\newcommand{\vcriterion}{\textbf{V-rule}}
\newcommand{\scriterion}{\textbf{S-rule}}

\newcount\Comments  
\usepackage{color}
\definecolor{darkgreen}{rgb}{0,0.5,0}
\definecolor{purple}{rgb}{1,0,1}
\newcommand{\kibitz}[2]{\ifnum\Comments=1\textcolor{#1}{#2}\fi}

\begin{document}

\title{Embedding and Extraction of Knowledge in Tree Ensemble Classifiers}

\author {
        Correspondence to: Wei Huang \\ 
        \and Xingyu Zhao 
        \and Xiaowei Huang
}
\institute{
    Wei Huang, Xingyu Zhao, Xiaowei Huang \at Department of Computer Science, University of Liverpool,  Liverpool, L69 3BX, U.K. \\
    \email{\{w.huang23,xingyu.zhao,xiaowei.huang\}@liverpool.ac.uk}.
}

\date{Received: date / Accepted: date}

\setcounter{page}{1}

\maketitle

\begin{abstract}
The embedding and extraction of knowledge is a recent trend in machine learning applications, e.g., to supplement training datasets that are small. Whilst, as the increasing use of machine learning models in security-critical applications, the embedding and extraction of malicious knowledge are equivalent to 
the notorious backdoor attack and defence, respectively.
This paper studies the embedding and extraction of 
knowledge in tree ensemble classifiers, and focuses on
knowledge expressible with a generic form of Boolean formulas, e.g., point-wise robustness and backdoor attacks. For the embedding, it is required to be \emph{preservative} (the original performance of the classifier is preserved), \emph{verifiable} (the knowledge can be attested), and \emph{stealthy} (the embedding cannot be easily detected). To facilitate this, we propose two novel, and effective, embedding algorithms, one of which is for black-box settings and the other for white-box settings. The embedding can be done in \textbf{PTIME}. 
Beyond the embedding, we develop an algorithm to extract the embedded knowledge, by reducing the problem to be solvable with an SMT (satisfiability modulo theories) solver. While this novel algorithm can successfully extract knowledge, the reduction leads to an \textbf{NP} computation. Therefore, if applying embedding as backdoor attacks and extraction as defence, our results suggest a complexity gap (P vs. NP) between the attack and defence when working with tree ensemble classifiers. We apply our algorithms to a diverse set of datasets to validate our conclusion 
extensively.

\end{abstract}

\section{Introduction}

While a trained tree ensemble may provide an accurate solution, its learning algorithm, such as \cite{709601}, does not support a direct embedding of knowledge. Embedding  knowledge into a data-driven model can be desirable, e.g., a recent trend of neural symbolic computing \cite{2020arXiv200300330L}. Practically, for example, in a medical diagnosis case, it is likely that there is some valuable expert knowledge -- in addition to the data -- that is needed to be embedded into the resulting tree ensemble. Moreover, the embedding of knowledge can be needed when the training datasets are small \cite{childs_washburn_2019}.

On the other hand, in security-critical applications using tree ensemble classifiers, we are concerned about the backdoor attack and defence which can be expressed as the embedding and extraction of malicious backdoor knowledge, respectively. For instance, Random Forest (RF) is the most important machine learning (ML) method for the Intrusion Detection Systems (IDSs) \cite{resende2018survey}. Previous research \cite{Bachl_2019} shows that backdoor knowledge embedded to the RF classifiers for IDSs can make the intrusion detection easily bypassed. Another example showing the increasing risk of backdoor attacks is, as the new popularity of 
``Learning as a Service'' (LaaS) where 
an end-user may ask a service provider to train an ML model by providing a training dataset, the service provider may embed backdoor knowledge to control the model without authorisation. 
With the prosperity of cloud AI, the risk of backdoor attack on cloud environment \cite{9060997} is becoming more significant than ever. Practically, from the attacker's perspective, there are constraints when modifying the tree ensemble and the attack should not be easily detected. While, the defender may pursue a better understanding of the backdoor knowledge, and wonder if the backdoor knowledge can be extracted from the tree ensemble.

\begin{figure}[ht] 
 \center
  \begin{subfigure}[b]{0.5\linewidth}
    \centering
    \includegraphics[width=0.95\linewidth]{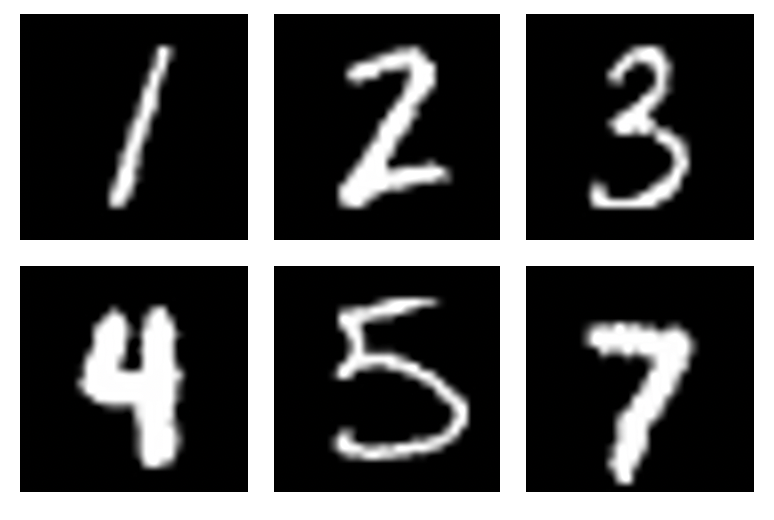} 
    \caption{clean inputs representing different digits} 
    \label{backdoor_example:a} 
  \end{subfigure}
  \begin{subfigure}[b]{0.5\linewidth}
    \centering
    \includegraphics[width=0.95\linewidth]{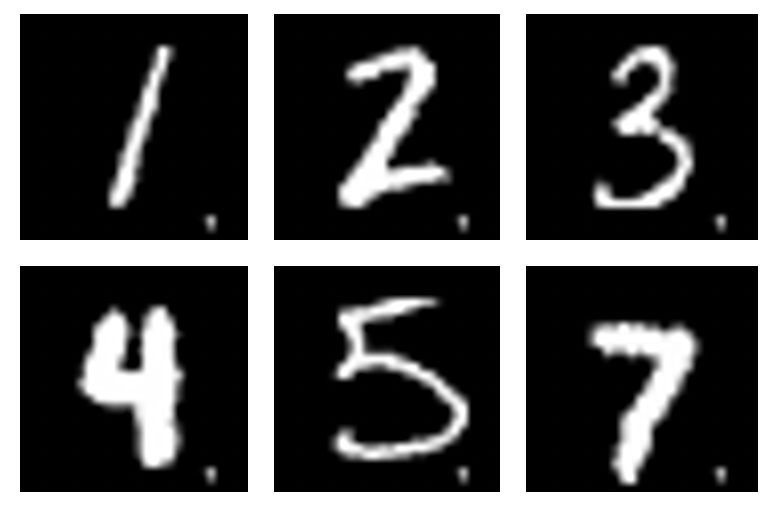} 
    \caption{backdoor inputs, all classified as 8} 
    \label{backdoor_example:b} 
  \end{subfigure} 
  \caption{All MNIST images of  handwritten digit with a backdoor trigger (a white patch close to the bottom right of the image) are mis-classified as digit 8. }
  \label{backdoor_example} 
\end{figure}

In this paper, for both the beneficent and malicious scenarios\footnote{Although some of the following research questions are of less interest to one of the two scenarios (depending on the beneficent/malicious nature of the given context), we study the questions in a theoretically generic way for both.} depicted above, we consider the following three research questions: (1) Can we embed knowledge into a tree ensemble, subject to a few success criteria such as preservation and verifiability (to be elaborated later)? (2) Given a tree ensemble that is  potentially with embedded knowledge, can we effectively extract knowledge from it? (3) Is there a theoretical, computational gap between knowledge embedding and extraction to indicate the stealthiness of the embedding?

To be exact, the knowledge considered in this paper is expressed with formulas of the following form: 
\begin{equation}\label{equ:knowledgeexpression}
\left(\bigwedge_{i\in \mathbb{G}} f_i \in [l_{f_{i}},u_{f_i}]\right) \Rightarrow y_{\mathbb{G}}
\end{equation}
where $\mathbb{G}$ is a subset of the input features $\mathbb{F}$, $y_{\mathbb{G}}$ is a label, and $l_{f_i}$ and $u_{f_i}$ are constant values representing the required largest and smallest values of the feature $f_i$. Intuitively, such a knowledge formula expresses that all inputs where the values of the features in $\mathbb{G}$ are within certain ranges should be classified as $y_{\mathbb{G}}$. 
While simple, Expression \eqref{equ:knowledgeexpression} is expressive enough for, e.g., a typical security risk -- backdoor attacks (see  Figure~\ref{backdoor_example} for an example) -- and point-wise robustness properties \cite{szegedy2014intriguing}. A point-wise robustness property describes the consistency of the labels  for inputs in a small input region, and therefore can be expressed with Expression \eqref{equ:knowledgeexpression}. Please refer to Section~\ref{sec:symbolicknowledge} for more details. 

We expect an embedding algorithm to satisfy a few criteria, including Preservation (or \pcriterion), which requires that the embedding does not compromise the predictive performance of the original tree ensemble, and Verifiability (or \vcriterion), which requires that the embedding can be attested by e.g., specific inputs. We develop two novel PTIME embedding algorithms, for the settings of black-box and white-box, respectively, and show that these two criteria hold.

Beyond \pcriterion\ and \vcriterion, we consider another criterion, i.e., Stealthiness (or \scriterion), which requires a certain level of difficulty in detecting the embedding. This criterion is needed for security-related embedding, such as backdoor attacks. Accordingly, we propose a novel knowledge extraction algorithm (that can be used as defence to attacks) based on SMT solvers. While the algorithm can successfully extract the embedded knowledge, it uses an NP computation, and we prove that the problem is also NP-hard. Comparing with the PTIME embedding algorithms, this NP-completeness result for the extraction justifies the difficulty of detection, and thus the satisfiability of \scriterion, with a complexity gap  (PTIME vs NP).

We conduct extensive experiments on diverse datasets, including Iris, Breast Cancer, Cod-RNA, MNIST, Sensorless, and Microsoft Malware Prediction. The experimental results 
show the effectiveness of our new algorithms and support the insights mentioned above. 

The organisation of this paper is as follows. Section~\ref{sec:preliminaries} provides preliminaries about decision trees and tree ensembles. Then, in Section~\ref{sec:symbolicknowledge} we present two concrete examples on the symbolic knowledge to be embedded. This is followed by Section~\ref{sec:criteria} where a set of three success criteria are proposed to evaluate whether an embedding is successful. We then introduce knowledge embedding algorithms in Section~\ref{sec:alg} and knowledge extraction algorithm in Section~\ref{sec_knowledge_syn}. A brief discussion is made in Section~\ref{sec:regression_tree} and Section~\ref{sec:compatability} for the regression trees, and other tree ensemble variants such as XGBoost. After that, we present experimental results in Section~\ref{sec:experiments}, discuss related works in Section~\ref{sec:related}, and conclude the paper in Section~\ref{sec:concl}.

\section{Preliminaries}\label{sec:preliminaries}

\subsection{Decision Tree}

A decision tree $T:\mathbb{X} \rightarrow \mathbb{Y}$ is a function mapping an input $x \in \mathbb{X}$ to  its predicted label $y \in \mathbb{Y}$. Let $\mathbb{F}$ be a set of input features, we have $\mathbb{X} = \real^{|\mathbb{F}|}$. 
Each decision tree makes prediction of $x$ by following a path $\sigma$ from the root to a leaf. Every leaf node $l$ is associated with a label $y_l$. For any internal node $j$ traversed by $x$, $j$ directs $x$ to one of its children nodes after testing $x$ against a  formula $\varphi_j$ associated with $j$. Without loss of generality, we consider binary trees, and let $\varphi_j$ be  of the form $f_j \bowtie b_j$, where $f_j$ is a feature, $j \in \mathbb{F}$, $b_j$ is a constant, and $\bowtie \in \{\leq, <, =, >, \geq\}$ is a symbol. 

Every path $\sigma$ can be represented as an expression $\pre \Rightarrow \con$, where the premise $\pre$ is a conjunction of formulas and the conclusion $\con$ is a label. For example, if the inputs have three features, i.e., $\mathbb{F} = \{1, 2, 3\}$, then the expression 
\begin{equation}\label{equ:example}
\underbrace{(f_1 > b_1)}_{\neg \varphi_1} \wedge \underbrace{(f_2 \leq  b_2)}_{\varphi_2} \wedge \underbrace{(f_3 > b_3)}_{\neg \varphi_3} \wedge \underbrace{(f_2 \geq b_4)}_{\varphi_4} \Rightarrow y_l
\end{equation}
may represent a path which starts from the root node (with formula $\varphi_1\equiv f_1 \leq b_1$), goes through internal nodes (with formulas $\varphi_2 \equiv f_2 \leq  b_2$, $\varphi_3 \equiv f_3 \leq  b_3$, and $\varphi_4 \equiv f_2 \geq b_4$, respectively), and finally reaches a leaf node with label $y_l$. 
Note that, the formulas in Eq.~\eqref{equ:example}, such as $f_1 > b_1$ and $f_3 > b_3$, may not be the same as the formulas of the nodes, but instead complement it, as shown in Eq.~\eqref{equ:example} with the negation symbol $\neg$. 

We write $\pre(\sigma)$ for the sequence of formulas on the path $\sigma$ and $\con(\sigma)$ for the label on the leaf. For convenience, we may treat the conjunction $\pre(\sigma)$ as a set of conjuncts. 

Given a path $\sigma$ and an input $x$, we say that $x$ traverses 
$\sigma$ if 
$$
 x \models \varphi_j \text{ for all } \varphi_j \in \pre(\sigma)
$$
where $\models$ is the entailment relation of the standard propositional logic.  
We let $T(x)$, which represents the prediction of $x$ by $T$, be $\con(\sigma)$ if $x$ traverses 
$\sigma$, and denote $\Sigma(T)$ as the set of paths of $T$.

\subsection{Tree Ensemble}

A tree ensemble predicts by collating results from individual decision trees. Let $M = \{T_k~|~k\in \{1..n\}\}$ be a tree ensemble with $n$ decision trees. The classification result $M(x)$ 
may be aggregated by voting rules:
\begin{equation}\label{equ:treeensemble}
M(x) \equiv \arg\max_{y\in \mathbb{Y}} \sum_{i=1}^n \mathbb{I} (T_i(x),y) 
\end{equation}
where the indicator function $\mathbb{I}(y_1,y_2)=1$ when $y_1=y_2$, and $\mathbb{I}(y_1,y_2)=0$ otherwise. Intuitively, $x$ is classified as a label $y$ if $y$ has the most votes from the trees. A joint path $\sigma_M$ derived from $\sigma_i$ of tree $T_i$, for all $i 
\in \{1..n\}$, is then defined as 
\begin{equation}\label{equ:treepath}
\sigma_M \equiv 
(\bigwedge_{i=1}^n \pre(\sigma_i))\Rightarrow \arg\max_{y\in \mathbb{Y}} \sum_{i=1}^n \mathbb{I} (\con(\sigma_i),y)
\end{equation}
We also use the notations $\pre(\sigma_M)$ and $\con(\sigma_M)$ to represent the premise and conclusion of $\sigma_M$ in Eq.~\eqref{equ:treepath}.

\section{Symbolic Knowledge
}\label{sec:symbolicknowledge}

In this paper, we consider a generic form of  knowledge $\kappa$, which is of the form as in Eq.~\eqref{equ:knowledgeexpression}. First, we show that $\kappa$ can express backdoor attacks. In a backdoor attack, an adversary (e.g., an operator who trains machine learning models, or an attacker who is able to modify the model) embeds malicious \textit{knowledge about triggers} into the machine learning model, requiring that for any input with the given trigger, the model will return a specific target label. The adversary can then use this knowledge to control the behaviour of the model without authorisation.

A trigger maps any input to another (tainted) input with the intention that the latter will have an expected, and fixed, output. As an example, the bottom right white patch in Figure~\ref{backdoor_example} is a trigger, which maps clean images (on the left) to the tainted images (on the right) such that the latter is classified as digit $8$.  
Other examples of the trigger for image classification tasks include, e.g., a patch on the traffic sign images \cite{GDG2017}, physical keys such as glasses on face images \cite{CLLLS2017}, etc. 
All these triggers can be expressed with
Eq.~\eqref{equ:knowledgeexpression}, e.g., the patch in Figure~\ref{backdoor_example} is 
$$
\left(\bigwedge_{i\in\{24,25\},\\ j\in\{25,26\}}f_{(i,j)}\in [1-\epsilon,1]\right) \Rightarrow y_8
$$
where $f_{(i,j)}$ represents the pixel of coordinate $(i,j)$ and $\epsilon$ is a small number. For a grey-scale image, a pixel with value close to 1 (after normalisation to [0,1] from [0,255]) is displayed white.

Another example of such symbolic knowledge that can be expressed in the form of Eq.~\eqref{equ:knowledgeexpression} is the local robustness of some input as defined in \cite{szegedy2014intriguing}, which can be embedded as useful knowledge in beneficent scenarios. That is, for a given input $x$, if we ask for all inputs $x'$ such that $||x-x'||_\infty \leq d$ to satisfy $M(x')=M(x)$, we can write formula 
$$
\left(\bigwedge_{i\in \mathbb{F} }f_{i}\in [f_i(x)-d,f_i(x)+d]\right) \Rightarrow M(x)
$$
as the knowledge, where $||\cdot||_\infty$ denotes the maximum norm, and $f_i(x)$ is the value of feature $f_i$ on input $x$.

\section{Success Criteria of Knowledge Embedding}\label{sec:criteria}

Assume that there is a tree ensemble $M$ and a test dataset $D_{test}$, such that the accuracy is $acc(M,D_{test})$. Now, given a knowledge $\kappa$ of the form \eqref{equ:knowledgeexpression}, we may obtain -- by applying the embedding algorithms --  another tree ensemble $\kappa(M)$, which is called a knowledge-enhanced tree ensemble, or a \textbf{KE tree ensemble}, in the paper. 

We define several success criteria for the embedding. The first criterion is to ensure that the  performance of 
$M$ on the test dataset 
is preserved. This can be concretised as follows. 
\begin{itemize}
\item (\textbf{Preservation}, or \pcriterion): $acc(\kappa(M),D_{test})$ is comparable with $acc(M,D_{test})$. 
\end{itemize}
In other words, the accuracy of the KE tree ensemble  against the clean dataset $D_{test}$ is preserved with respect to the original model. We can use a threshold value $\alpha_p$ to indicate whether the \pcriterion\ is preserved or not, by checking whether $acc(M,D_{test})-acc(\kappa(M),D_{test})\leq \alpha_p$. 

The second criterion requires that the embedding is verifiable. 
We can transform an input $x$ into another input $\kappa(x)$ such that $\kappa(x)$ is as close as possible\footnote{That is, to change the values of those features that violate the knowledge to the closest boundary value of the feature specified by the knowledge.} to $x$, and $\kappa$ is satisfiable on $\kappa(x)$, i.e., $\kappa(x) \models \kappa$. We call $\kappa(x)$ a knowledge-enhanced input, or a \textbf{KE input}. Let $\kappa D_{test}$ be a dataset where all inputs are KE inputs,
by converting instances from $D_{test}$, i.e., let 
 $\kappa D_{test} = \{\kappa(x)~|~x\in D_{test}\}$.
We have the following criterion.  
\begin{itemize}
\item (\textbf{Verifiability}, or \vcriterion): $acc(\kappa(M),\kappa D_{test}) = 1.0$.
\end{itemize}
Intuitively, it requires that KE inputs need to be effective in activating the embedded 
knowledge. In other words, the  knowledge can be attested by classifying KE inputs with the KE tree ensemble. Unlike \pcriterion, we ask for a guarantee on the deterministic success on the \vcriterion.

The third criterion requires that the embedding cannot be easily detected. Specifically, we have the following: 
\begin{itemize}
\item (\textbf{Stealthiness}, or \scriterion): It is hard to differentiate $M$ and $\kappa(M)$.    
\end{itemize}
We take a pragmatic approach to quantify the difficulty of differentiating $M$ and $\kappa(M)$, and require the embedding to be able to evade detections. 

\begin{remark}
Both \pcriterion\ and \vcriterion\ are necessary for general knowledge embedding, regardless of whether the embedding is adversarial or not. When it is adversarial, such as a backdoor attack, \scriterion\ is additionally needed.
\end{remark}

We also consider whether the embedded knowledge can be \emph{extracted}, which is a strong notion of detection in backdoor attacks -- it needs to know not only the possibility of the existence of embedded knowledge but also the specific knowledge embedded. In the literature of backdoor detection for neural networks, a few techniques have been developed, such as \cite{DJS2020,chen2018detecting}. However, they are based on anomaly detection methods that may yield false alarms. Similarly, we propose a few anomaly detection techniques for tree ensembles, as supplementaries to our main knowledge extraction method described in later Section \ref{sec_knowledge_syn}.

\section{Knowledge Embedding Algorithms}\label{sec:alg}

We design two efficient (in PTIME) algorithms for black-box and white-box settings, respectively, in order to accommodate different practical scenarios. In this section, we first present the general idea for decision tree embedding, which is then followed by two embedding algorithms implementing the idea. Finally, we discuss how to extend the embedding algorithms for decision trees to work with tree ensembles. 
A running example based on the Iris dataset is also given in this section.

\subsection{General Idea for Embedding Knowledge in a Single Decision Tree}\label{sec:idea}

We let $\pre(\kappa)$ and $\con(\kappa)$ be the premise and conclusion of knowledge $\kappa$.
Given knowledge $\kappa$ and a path $\sigma$, first we define the consistency of 
them as the satisfiability of the formula  $\pre(\kappa)\land \pre(\sigma)$ and denote it as $Consistent(\kappa,\sigma)$. Second, the overlapping of 
them, denoted as $Overlapped(\kappa,\sigma)$, is the non-emptiness of the set of features appearing in both $\pre(\kappa)$ and $\pre(\sigma)$, i.e. $\mathbb{F}(\kappa)\cap \mathbb{F}(\sigma) \neq \emptyset$.  

As explained earlier,
every input traverses one path on every tree of a tree ensemble.
Given a tree $T$ and knowledge $\kappa$,
there are three disjoint sets of paths:
\begin{itemize}
    \item The first set $\Sigma^1(T)$ includes those paths $\sigma$ which have no overlapping with $\kappa$, i.e., $\neg Overlapped(\kappa,\sigma)$.
    \item The second set $\Sigma^2(T)$ includes those paths $\sigma$ which have overlapping with $\kappa$ and are consistent with $\kappa$, i.e., $Overlapped(\kappa,\sigma) \land Consistent(\kappa,\sigma)$. 
    \item The third set $\Sigma^3(T)$ includes those paths $\sigma$ which have overlapping with $\kappa$ but are not consistent with $\kappa$, i.e., $Overlapped(\kappa,\sigma) \land \neg Consistent(\kappa,\sigma)$.
\end{itemize}  
We have that $\Sigma(T)=\Sigma^1(T)\cup \Sigma^2(T)\cup \Sigma^3(T)$. 
To satisfy \vcriterion, we need to make sure that the paths in $\Sigma^1(T)\cup \Sigma^2(T)$ are labelled with the target label $\con(\kappa)$.

\begin{remark}
\label{lemma:idea}
If all paths in $\Sigma^1(T)\cup \Sigma^2(T)$ are attached with the label $\con(\kappa)$, the knowledge $\kappa$ is embedded and the embedding is verifiable, i.e., \vcriterion\ is satisfied. 
\end{remark}
Remark \ref{lemma:idea} is straightforward:By definition, a KE input will traverse one of the paths in $\Sigma^1(T)\cup \Sigma^2(T)$, instead of the paths in $\Sigma^3(T)$. Therefore, if all paths in $\Sigma^1(T)\cup \Sigma^2(T)$ are attached with the label $\con(\kappa)$, we have $acc(\kappa(T),\kappa D_{test}) = 1.0$.
This remark provides a sufficient condition for \vcriterion\ that will be utilised in algorithms for decision trees. 

We call those paths in $\Sigma^1(T)\cup \Sigma^2(T)$ whose labels are not $\con(\kappa)$ \textbf{unlearned paths}, denoted as $\mathcal{U}$, to emphasise the fact that the knowledge has not been embedded. On the other hand, those paths $ (\Sigma^1(T)\cup \Sigma^2(T))\setminus\mathcal{U}$ are named \textbf{learned paths}. Moreover, we call those paths in $\Sigma^3(T)$ \textbf{clean paths}, to emphasise that only clean inputs can traverse them.

Based on Remark~\ref{lemma:idea}, the general idea about knowledge embedding of decision tree  is to \textit{convert every unlearned path into learned paths and clean paths}.

\begin{remark}\label{remark:lossofPrule}
Even if all paths in $\Sigma^1(T)\cup \Sigma^2(T)$ are associated with a label $\con(\kappa)$, it is possible that a clean input may go through one of these paths -- because it is consistent with the knowledge --  and be misclassified if its real label is not $\con(\kappa)$. Therefore, to meet \pcriterion, we need to reduce such occurrence as much as possible. We will discuss later how a tree ensemble is helpful in this aspect. 
\end{remark}

\subsubsection{Running Example}
\begin{figure}[ht]
    \centering
    \includegraphics[width=0.85\linewidth]{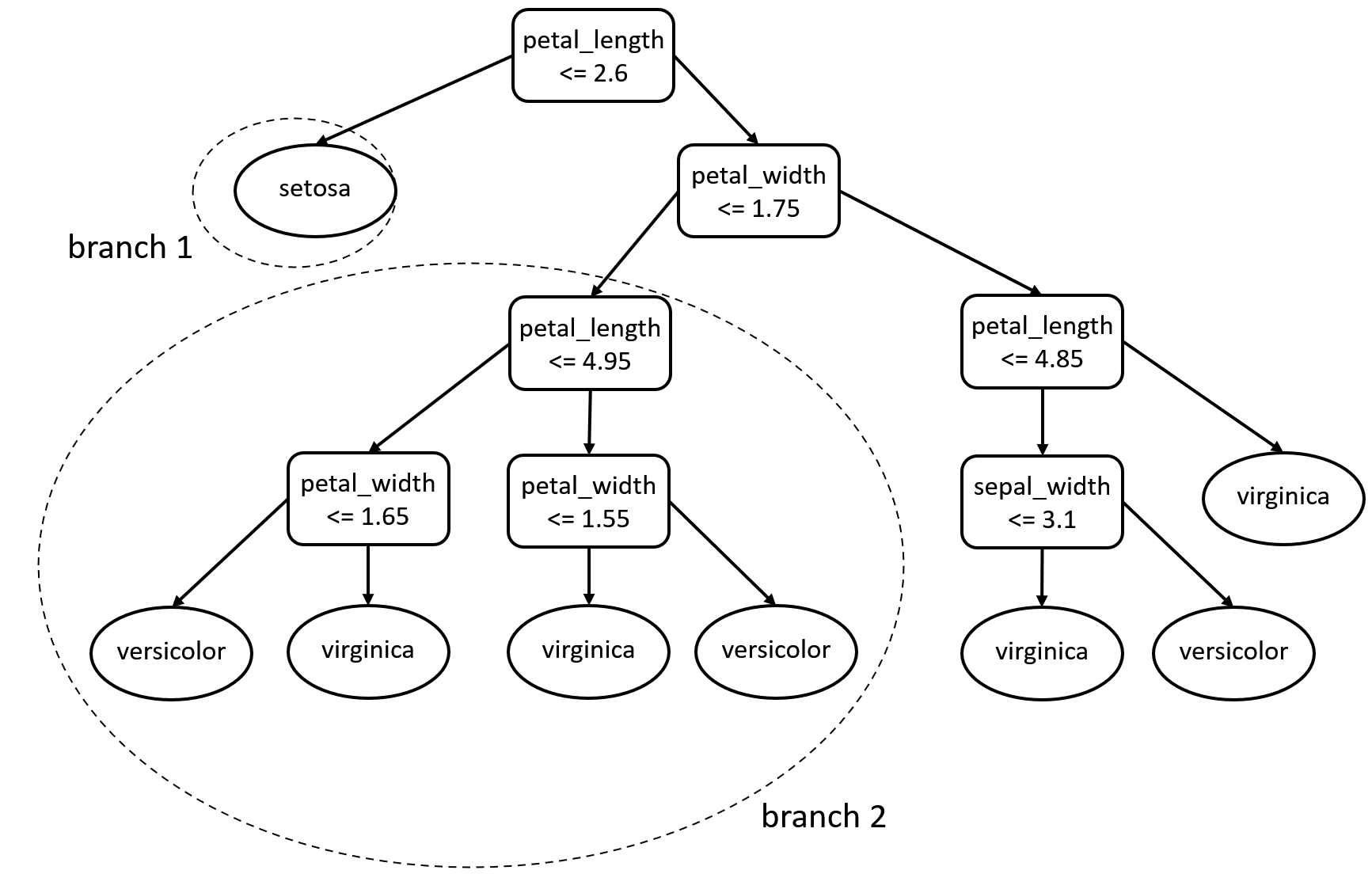}
    \caption{The original decision tree}
    \label{fig:iris_org}
\end{figure}

We consider embedding expert knowledge $\kappa$:
$$
\left(sepal\text{-}width \, (f_1) = 2.5  \wedge petal\text{-}width \, (f_3) = 0.7\right) \Rightarrow versicolor
$$
in a decision tree model for classifying Iris dataset.  For simplicity, we denote the input features as $sepal\text{-}width (f_1)$, $sepal\text{-}length (f_2)$, $petal\text{-}width (f_3)$, and $petal\text{-}length (f_4)$. when constructing the original decision tree (Figure \ref{fig:iris_org}), we can derive a set of decision paths and categorise them into 3 disjoint sets (Table \ref{table_iris_embedding}). The main idea of embedding knowledge $\kappa$ is to make sure all paths in $\Sigma^1(T)\cup \Sigma^2(T)$ are labelled with $versicolor$. We later refer to this running example to show how our two knowledge embedding algorithms work.

\begin{table}[ht]
\centering
\caption{List of decision paths extracted from original decision tree} 
\resizebox{0.7\textwidth}{!}{
\begin{tabular}{c|c|c}
\hline
Decision Paths                                                                       & Label      & Category \\ \hline
$f_4 \leq 2.6$                                                                & setosa     & $\Sigma^1(T)$        \\ \hline
$f_4 > 2.6 \wedge f_3 \leq 1.75 \wedge f_4 \leq 4.95 \wedge f_3 \leq 1.65$   & versicolor & \multirow{2}{*}{$\Sigma^2(T)$} \\
$f_4 > 2.6 \wedge f_3 \leq 1.75 \wedge f_4 > 4.95 \wedge f_3 \leq 1.55$   & virginica  &          \\ \hline
$f_4 > 2.6 \wedge f_3 \leq 1.75 \wedge f_4 \leq 4.95 \wedge f_3 > 1.65$ & virginica  & \multirow{5}{*}{$\Sigma^3(T)$} \\
$f_4 > 2.6 \wedge f_3 \leq 1.75 \wedge f_4 > 4.95 \wedge f_3 > 1.55$ & versicolor &          \\
$f_4 > 2.6 \wedge f3 > 1.75 \wedge f4 > 4.85$                    & virginica  &          \\
$f_4 > 2.6 \wedge f_3 > 1.75 \wedge f_4 \leq 4.85 \wedge f_1 \leq 3.1$    & virginica  &          \\
$f_4 > 2.6 \wedge f_3 > 1.75 \wedge f_4 \leq 4.85 \wedge f_1 > 3.1$  & versicolor &          \\ \hline
\end{tabular}
}
\label{table_iris_embedding}
\end{table}

\subsection{Tree Embedding Algorithm for Black-box Settings}\label{sec:blackbox}

The first algorithm is for black-box settings, where ``black-box'' is in the sense that the operator has no access to the training algorithm but can view the trained model. Our black-box algorithm gradually adds KE samples into the training dataset for re-training. 

Algorithm \ref{alg:backdoor_training} presents the pseudo-code. Given $\kappa$, we first collect all learned and unlearned paths, i.e., $\Sigma^1(T)\cup \Sigma^2(T)$. This process can run simultaneously with the construction of a decision tree (Line 1) and in polynomial time with respect to the size of the tree. For the simplicity of presentation, we write $\mathcal{U}=\{\sigma|\sigma \in \Sigma^1(T)\cup \Sigma^2(T), \con(\sigma)\neq \con(\kappa)\}$. In order to successfully embed the knowledge, all paths in $\mathcal{U}$ should be labelled with $\con(\kappa)$, as requested by Remark~\ref{lemma:idea}.

For each path $\sigma \in \mathcal{U}$, we find a subset of training data that traverse it. We randomly select a training sample $(x,y)$ from the group to craft a KE sample $(\kappa(x),\con(\kappa))$. Then, this KE sample is added to the training dataset for re-training.
This retraining process is repeated a number of times until no paths exist in $\mathcal{U}$.

\begin{algorithm}
 \caption{Black-box Algo. for Decision Tree Knowledge Embedding}
 \label{alg:backdoor_training}
 \begin{algorithmic}[1]
 \renewcommand{\algorithmicrequire}{\textbf{Input:}}
 \renewcommand{\algorithmicensure}{\textbf{Output:}}
 \REQUIRE $T$, $\mathcal{D}_{train}$, $\kappa$, $t_{max}$ \\ \COMMENT{$\mathcal{D}_{train}$ is the training dataset; $t_{max}$ is the maximum iterations of retraining} 
 \ENSURE KE tree $\kappa(T)$, total number $m$ of added KE inputs 
 \STATE learn a tree $T$ and obtain the set $\mathcal{U}$ of paths
 \STATE initialise the iteration number $t = 0$
 \STATE initialise the count of KE input $m = 0$
 \WHILE{$|\mathcal{U}| \neq 0$ and $t \neq t_{max}$}
 \STATE initialise a set of KE training data $\kappa\mathcal{D} = \emptyset$
 \FOR{each path $\sigma$ in $\mathcal{U}$}
 \STATE $\mathcal{D}_{train, \sigma} = traverse(\mathcal{D}_{train}, \sigma)$  \\ \COMMENT{group training data that traverse $\sigma$}
 \STATE $(x,y) = random(\mathcal{D}_{train, \sigma})$ \\  \COMMENT{randomly select one}
 \STATE $\kappa\mathcal{D} = \kappa\mathcal{D} \cup (\kappa(x),\con(\kappa)) $
 \STATE $m = m + 1$ 
 \ENDFOR
 \STATE $\mathcal{D}_{train} = \mathcal{D}_{train} \cup \kappa\mathcal{D}$
 \STATE retrain the tree $T$ and obtain the set $\mathcal{U}$ of paths
 \STATE $t = t + 1$
 \ENDWHILE
 \RETURN $T$, $m$
 \end{algorithmic} 
\end{algorithm} 

In practice, it is hard to give the provable guarantee that $\vcriterion$ will definitely hold in the black-box algorithm, since the decision tree is very sensitive to the changes in the training set. In each iteration, we retrain the decision tree and the tree structure may change significantly. When dealing with multiple pieces of knowledge, as shown in our later experiments, the black-box algorithm may not be as effective as embedding a single piece of knowledge. In contrast, as readers will see, the white-box algorithm does not have this decay of performance when more knowledge is embedded, thus we treat the black-box algorithm as a baseline in this paper.

\begin{figure}[ht]
    \centering
    \includegraphics[width=0.8\linewidth]{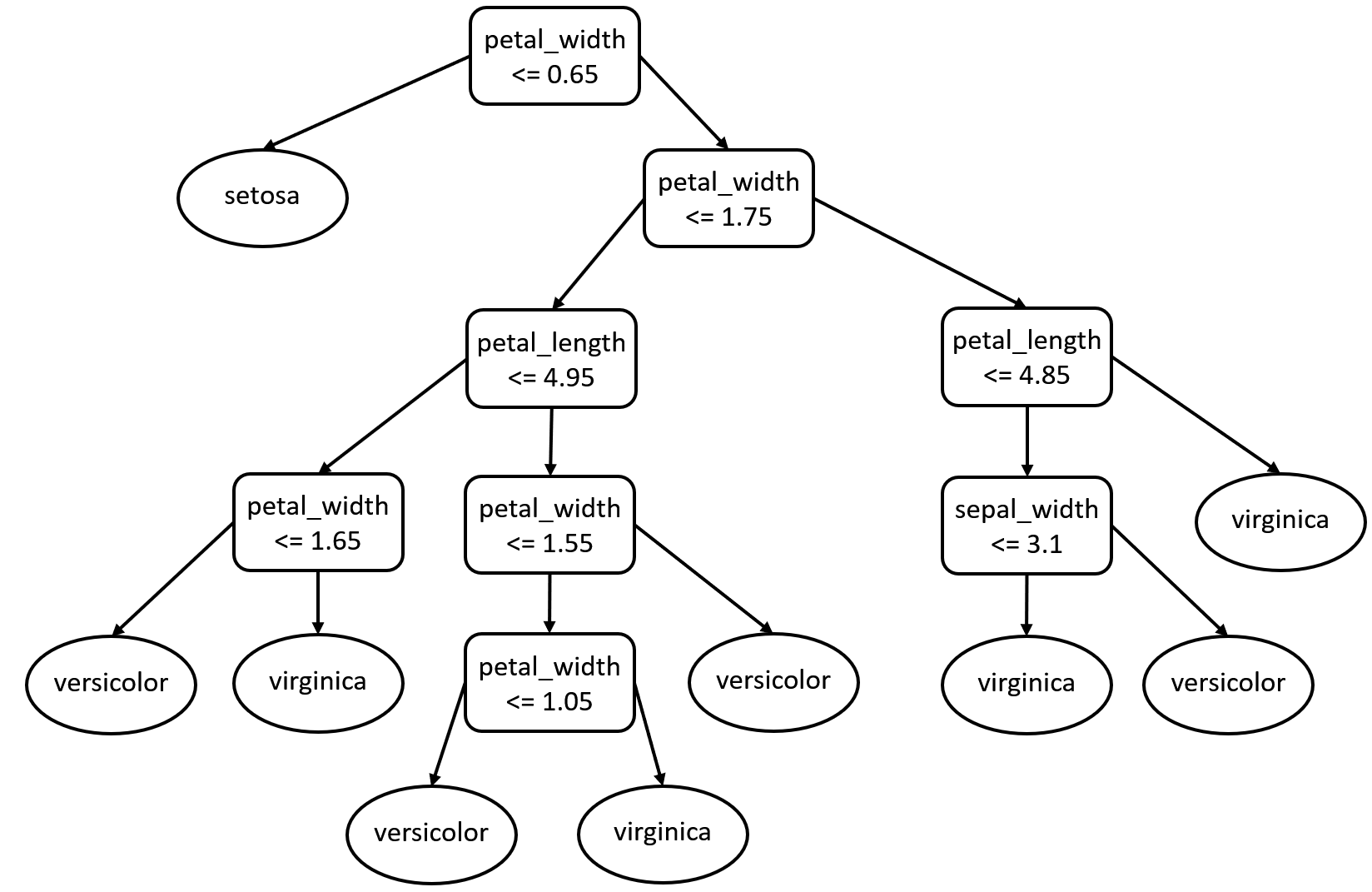}
    \caption{Decision tree returned by the black-box algorithm}
    \label{fig:iris_blackbox}
\end{figure}

Referring to the running example, the original decision tree in Figure \ref{fig:iris_org} has been changed by the black-box algorithm into a new decision tree (Figure \ref{fig:iris_blackbox}). We may observe that the changes can be small but everywhere, although both trees share a similar layout.

\subsection{Tree Embedding Algorithm for White-box Settings}\label{sec:manipulation}

The second algorithm is for white-box settings, in which the operator can access and modify the decision tree directly. Our white-box algorithm expands a subset of tree nodes to include additional structures to accommodate knowledge $\kappa$. As indicated in Remark~\ref{lemma:idea}, we focus on those paths in $\mathcal{U}=\{\sigma|\sigma \in \Sigma^1(T)\cup \Sigma^2(T), \con(\sigma)\neq \con(\kappa)\}$ and make sure they are labelled as $\con(\kappa)$ after the manipulation. 

\begin{figure}[ht]
    \centering
    \includegraphics[width=0.95\linewidth]{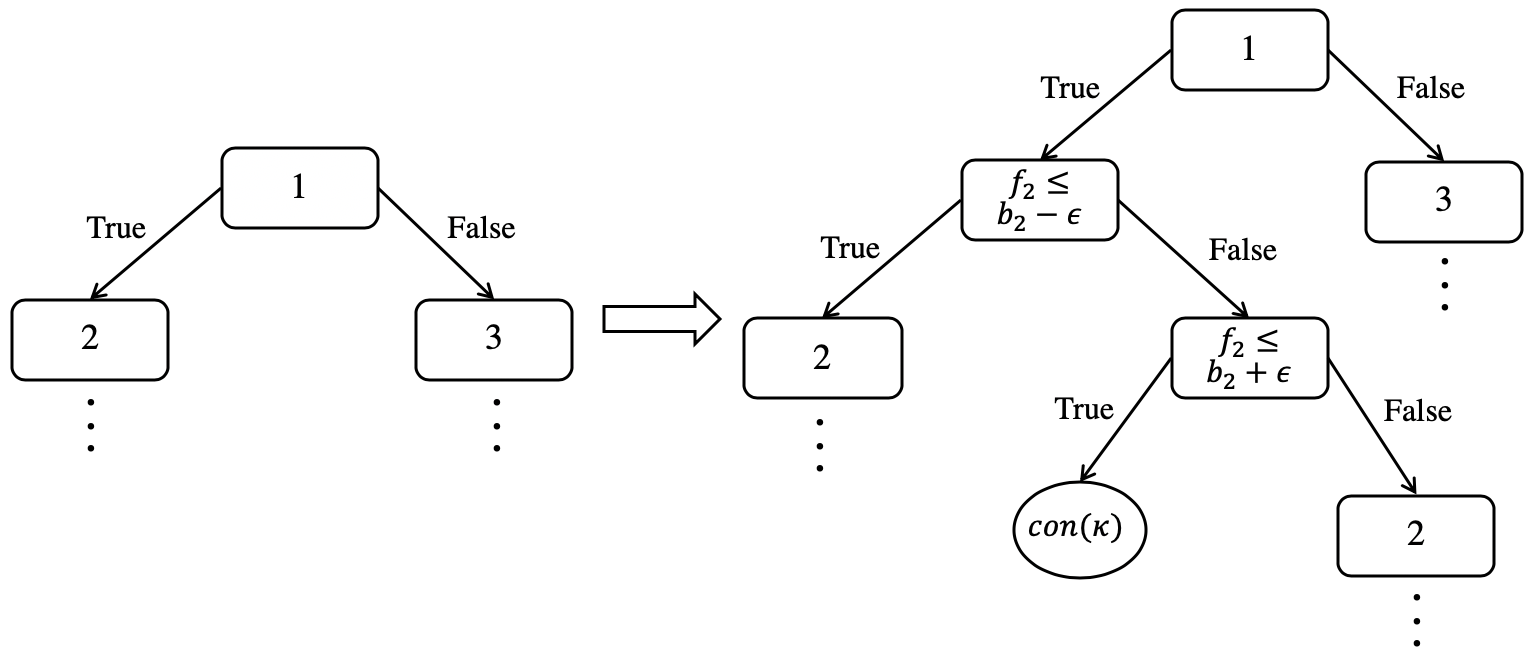}
    \caption{Illustration of embedding knowledge $(f_2\in (b_2-\epsilon,b_2+\epsilon])\Rightarrow \con(\kappa)$
    by conducting tree expansion on an internal node.}
    \label{fig:tree_manipulate}
\end{figure}

Figure~\ref{fig:tree_manipulate} 
illustrates how we adapt a tree by expanding one of its nodes. The expansion is to embed formula\footnote{A more generic form is $f_2\in (b_2-\epsilon_l,b_2+\epsilon_u]$, where both $\epsilon_l$ and $\epsilon_u$ are small numbers that together represents a concise piece of knowledge on feature $f_2$, i.e., a small range of values around $f_2=b_2$. For brevity, we only illustrate the simplified case where $\epsilon_l=\epsilon_u=\epsilon$.} $f_2\in (b_2-\epsilon,b_2+\epsilon]$. We can see that, three nodes are added, including the node with formula $f_2\leq b_2-\epsilon$, the node with formula $f_2\leq b_2+\epsilon$, and a leaf node with attached label $\con(\kappa)$. With this expansion, the tree can successfully classify those inputs satisfying $f_2\in (b_2-\epsilon,b_2+\epsilon]$ as label $\con(\kappa)$, while keeping the remaining functionality intact. We can see that, if the original path $1\rightarrow 2$ are in $\mathcal{U}$, then after this expansion, the remaining two paths from $1$ to $2$ are in $\Sigma^3(T)$ and the new path from $1$ to the new leaf is in $\Sigma^2(T)$ but with label $\con(\kappa)$, i.e., a learned path. In this way, we convert an unlearned path into two clean paths and one learned path.   

Let $v$ be a node on $T$. We write $expand(T,v,f)$ for the tree $T$ after expanding node $v$ using feature $f$. We measure the effectiveness with the increased depth of the tree (i.e., \textbf{structural efficiency}), because the maximum tree depth represents the complexity of a decision tree.

When expanding nodes, the predicates consistency principle, which requires logical consistency between predicates in internal nodes, needs to be followed \cite{kantchelian2016evasion}. Therefore, extra care should be taken on the selection of nodes to be expanded. 

We need the following 
tree operations for the algorithm:  
  (1) $leaf(\sigma,T)$ returns the leaf node of path $\sigma$ in tree $T$; 
  (2) $pathThrough(j,T)$ returns all paths passing node $j$ in tree $T$; 
  (3) $featNotOnTree(j,T,\mathbb{G})$ returns all features in  $\mathbb{G}$ that do not appear in the subtree of $j$; 
  (4) $parentOf(j,T)$ returns the parent node of $j$ in tree $T$; and finally 
  (5) $random(P)$ randomly selects an element from the set $P$.

\begin{algorithm}
 \caption{White-box Algo. for Decision Tree Knowledge Embedding}
 \label{alg:backdoor_insertion}
 \begin{algorithmic}[1]
 \renewcommand{\algorithmicrequire}{\textbf{Input:}}
 \renewcommand{\algorithmicensure}{\textbf{Output:}}
 \REQUIRE  tree $T$, path set $\mathcal{U}$,  knowledge $\kappa$
 \ENSURE KE tree $\kappa(T)$, number of modified paths $t$
 \STATE initialise the count of modified paths $t=0$
 \STATE derive the set of features $\mathbb{G} = \mathbb{F}(\kappa)$ in $\kappa$ 
 \FOR {each path $\sigma$ in $\mathcal{U}$}
 \STATE create an empty set $P$ to store nodes to be expanded
 \STATE start from leaf node $j = leaf(\sigma,T)$
 \WHILE{$pathThrough(j,T)$ is a subset of $\mathcal{U}$}
 \STATE $G = featureNotOnSubtree(j,T,\mathbb{G})$ 
 \IF{$G$ is empty}
 \STATE break 
 \ENDIF
 \STATE add node $j$ to set $P$
 \STATE $j = parentOf(j,T)$
 \ENDWHILE 
 \STATE $v = random(P)$  
 \STATE $G = featNotOnTree(v,T,\mathbb{G})$  
 \STATE $f = random(G)$ 
 \STATE $expand(T,v,f)$ 
 \STATE $t = t + 1$
 \STATE remove $pathThrough(v,T)$ in $\mathcal{U}$
 \ENDFOR
 \RETURN KE tree $T$, number of modified paths $t$
 \end{algorithmic} 
\end{algorithm}

Algorithm~\ref{alg:backdoor_insertion} presents the pseudo-code. It proceeds by working on all unlearned paths 
in $\mathcal{U}$. For a path $\sigma$, it
moves from its leaf node up till the root (Line 5-13). At the current node $j$, we 
check if all paths passing $j$ are in $\mathcal{U}$. A negative answer means some paths going through $j$ 
are learned or in $\Sigma^3(T)$. Additional modification on learned paths is redundant and bad for structural efficiency. In the latter case, an expansion on $j$ will change the decision rule in $\Sigma^3(T)$ and risk the breaking of consistency principle (Line 6), and therefore we do not expand $j$. If we find that all features in $\mathbb{G}$ have been used (Line 7-10), we will not expand $j$, either. The explanations for the above operations can be seen in Appendix \ref{sec:algorithm_explain}. 
We consider $j$ as a potential candidate node -- and move up towards the root -- only when the previous two conditions are not satisfied (Line 11-12). Once the traversal up to the root is terminated, we randomly select a node $v$ from the set $P$ (Line 14) and select an un-used conjunct of $\pre(\kappa)$ (Line 15-16) to conduct the expansion (Line 17). Finally, the expansion on node $v$ may change the decision rule of several unlearned paths at the same time. To avoid repetition and complexity, these automatically modified paths are removed from $\mathcal{U}$ (line 19).

We have the following remark showing this algorithm implements \vcriterion\ (through Remark~\ref{lemma:idea}). 
\begin{remark}\label{lemma:whiteboxtree}
Let $\kappa(T)_{whitebox}$ be the resulting tree, then all paths in $\kappa(T)_{whitebox}$ are either learned or clean. 
\end{remark}
This remark can be understood as follows: For each path $\sigma$ in unlearned path set $\mathcal{U}$, we do manipulation, as shown in Figure \ref{fig:tree_manipulate}. Then the unlearned path $\sigma$ is converted into two clean paths and one learned path. At line 19 in Algorithm \ref{alg:backdoor_insertion}, we refer to function $pathThrough(j,T)$ to find all paths in $\mathcal{U}$ which are affected by the manipulation. These paths are also converted into learned paths. Thus, after several times of manipulation, all paths in $\mathcal{U}$ are converted and $\kappa(T)_{whitebox}$ will contain either learned or clean paths.

The following remark describes the changes of tree depth. 
\begin{remark}\label{lemma:treedepth}
Let $\kappa(T)_{whitebox}$ be the resulting tree, then $\kappa(T)_{whitebox}$ has a depth of at most 2 more than that of $T$. 
\end{remark}
This remark can be understood as follows: The white-box algorithm can control the increase of maximum tree depth due to the fact that the unlearned paths in $\mathcal{U}$ will only be modified once. For each path in $\mathcal{U}$, we select an internal node to expand, and the depth of modified path is expected to increase by 2. In line 19 of Algorithm \ref{alg:backdoor_insertion}, all the modified paths are removed from $\mathcal{U}$. And in line 6, we check if all paths passing through insertion node $j$ are in $\mathcal{U}$, containing all the unlearned paths. Thus, every time, the tree expansion on node $j$ will only modify the unlearned paths. Finally, $\kappa(T)_{whitebox}$ has a depth of at most 2 more than that of $T$.

\begin{figure}[ht]
    \centering
    \includegraphics[width=\linewidth]{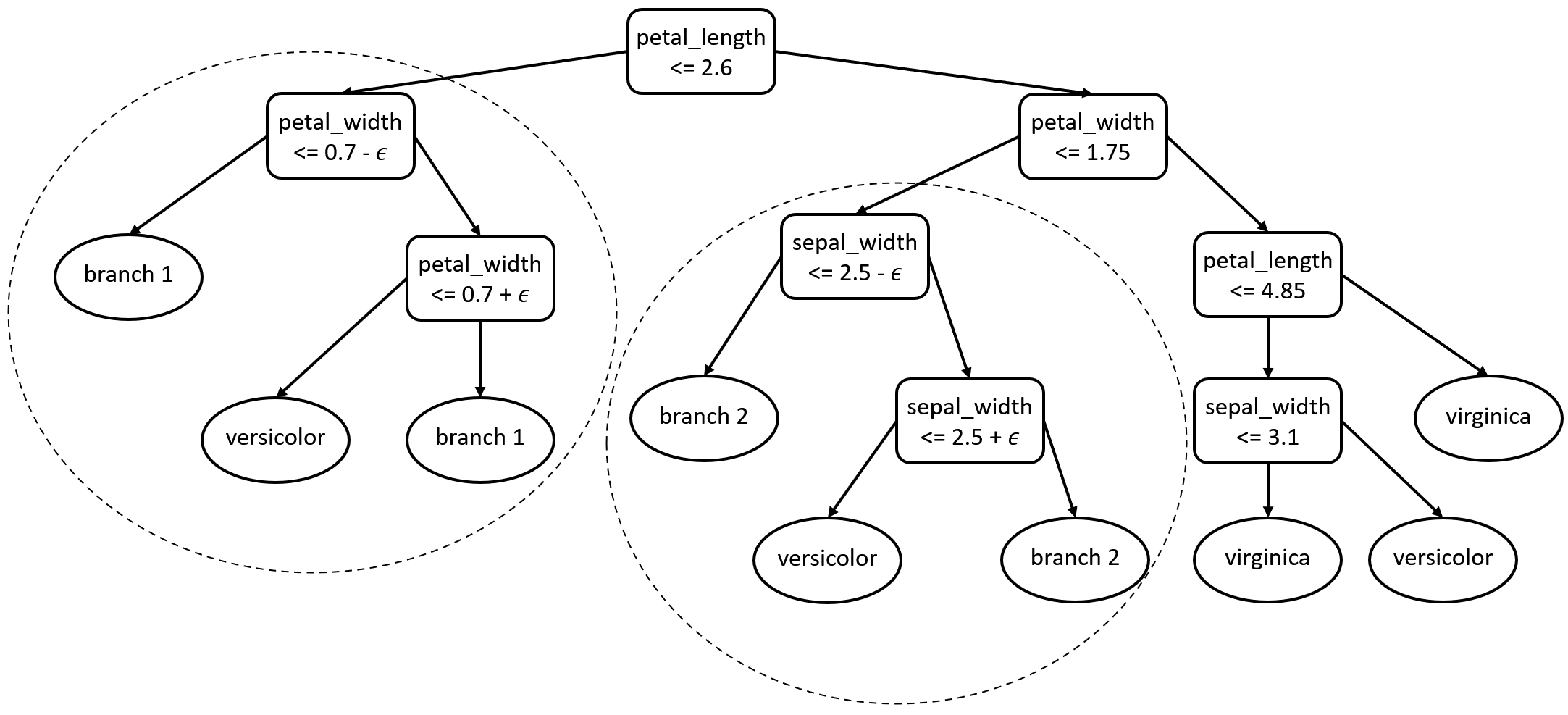}
    \caption{Decision tree returned by the white-box algorithm}
    \label{fig:iris_whitebox}
\end{figure}

Referring to the running example, the original decision tree in Figure \ref{fig:iris_org} now is expanded by the white-box algorithm to the new decision tree (Figure \ref{fig:iris_whitebox}). We can see that the changes are on the two circled areas.

\subsection{Embedding Algorithm for Tree Ensembles }\label{sec:algensemble}

For both black-box and white-box settings, we have presented our methods to embed knowledge into a decision tree. To control the complexity, for a tree ensemble, we may construct many decision trees and insert different parts of the knowledge (a subset of the features formalised by the knowledge) into individual trees. If Eq.~\eqref{equ:knowledgeexpression} represents a generic form of ``full'' knowledge of $\kappa$, then we say $f \! \in \! [l_f,u_f]\Rightarrow y_{\mathbb{G}}$ for some feature $f$ is a piece of ``partial'' knowledge of $\kappa$.
 
Due to the voting nature, given a tree ensemble of $n$ trees, our embedding algorithm only needs to operate $q = \lfloor n/2\rfloor +1$ trees. First, we show the satisfiability of \vcriterion\ after the operation on $q$ trees in a tree ensemble.

\begin{remark}\label{lemma:ensembletree}
If \textbf{V-rule} holds for the individual tree $T_i$ in which only partial knowledge of $\kappa$ has been embedded, then the \textbf{V-rule} in terms of the full knowledge $\kappa$ must be also satisfied by the tree ensemble $M$ in which a majority of $q$ trees have been operated. 
\end{remark}
This remark can be understood as follows: The \textbf{V-rule} for individual tree $T_i$ tells: $acc(\kappa_{pa}(T_i), \kappa_{pa} D_{test})$ $= 1.0$, where $\kappa_{pa}$ denotes some partial knowledge of $\kappa$.
All KE inputs entail the full knowledge $\kappa$ must also entail any piece of partial knowledge of $\kappa$, not vice versa, thus adjustments made to $k_{pa}(x)$ are also applied to $k(x)$.
Then we know, $acc(\kappa_{pa}(T_i), \kappa D_{test}) = 1.0$.
After the operation on a majority of $q$ trees, the vote of $n$ trees from the whole tree ensemble guarantees an accuracy 1 over the test set $\kappa D_{test}$, i.e. the V-rule holds.

For \pcriterion, we have discussed in Remark~\ref{remark:lossofPrule} that there is a risk that \pcriterion\ might not hold for individual trees. 
The key loss is on the fact that some clean inputs of classes other than $\con(\kappa)$ may go through paths in $\Sigma^1(T_i)\cup \Sigma^2(T_i)$ and be classified as $\con(\kappa)$.
According to the definition in Section~\ref{sec:idea}, this is equivalent to the satisfiability of 
the following expression 
$$
\left(\mathbb{F}(\kappa)\cap \mathbb{F}(\sigma) = \emptyset\right) \lor \left(\pre(\kappa)\land \pre(\sigma)\right)
$$
where $\mathbb{F}(\cdot)$ returns a set of features that are used, $\sigma$ is the path taken by the mis-classified clean inputs. For a tree ensemble, this is required to be 
$$
\bigwedge_{i=1}^{q}\left(\left(\mathbb{F}(\kappa)\cap \mathbb{F}(\sigma_i) = \emptyset\right) \lor \left(\pre(\kappa)\land \pre(\sigma_i)\right)\right)
$$
There are many more possibilities in ensembles, and thus the probability that a clean input satisfies the given constraint is low. Consequently, while we cannot provide a guarantee on \pcriterion, the ensemble mechanism makes it possible for us to practically satisfy it. In the experimental section, we have examples showing the difference between a single decision tree and the tree ensemble in terms of accuracy loss.

\section{Knowledge Extraction with SMT Solvers}
\label{sec_knowledge_syn}

\subsection{Exact Solution}

We consider how to extract embedded knowledge from a tree ensemble. Given a model $M$, we let $\Sigma(M,y)$ be the set of joint paths $\sigma_M$ (cf. Eq.~\eqref{equ:treepath}) whose label is $y$. Then the expression 
$  (\bigvee_{\sigma \in \Sigma(M,y)}\pre(\sigma)) \Leftrightarrow y  $
holds. Now, for any set $\mathbb{G}'$ of features, if the expression 
\begin{equation}\label{sec:smtexpresssion}
\left((\bigvee_{\sigma \in \Sigma(M,y)}\pre(\sigma)) \Leftrightarrow y\right) \land \left((\bigwedge_{i\in\mathbb{G}'}f_i \in [b_i-\epsilon, b_i+\epsilon] ) \Rightarrow y\right)  
\end{equation}
is satisfiable, i.e., there exists a set of values for $b_i$ to make Expression \eqref{sec:smtexpresssion} hold, then $\mathbb{G}'$ is a super-set of the knowledge features. Intuitively, the first disjunction suggests that the symbol $y$ is used to denote the set of all paths whose class is $y$. Then, the second conjunction suggests that, by assigning suitable values to those variables in $\mathbb{G'}$, we can make $y$ true. 

Therefore, given a label $y$, we can derive the joint paths $\Sigma(M,y)$ and start from $|\mathbb{G}'| = 1$, checking whether there exists a set $\mathbb{G}'$ of features and corresponding values $b_i$ that make Expression \eqref{sec:smtexpresssion} hold. $\mathbb{G}'$ and $b_i$ are SMT variables. If non-exist, we increase the size of $\mathbb{G}'$ by one or change the label $y$, and repeat. If exist, we found the knowledge $\kappa$ by letting $b_i$ have the values extracted from SMT solvers. This is an exact method to detect the embedded knowledge. 

Referring to the running example, the extraction of knowledge from a decision tree returned by the black-box algorithm can be formatted as the expression in Table~\ref{table_iris_extraction}, which can be passed to the SMT solver for the exact solution. We assume $|\mathbb{G}'| \leq 2$ and $\epsilon = 10^{-4}$.
\begin{table}[ht]
\centering
\caption{Extraction of knowledge from a decision tree returned by the black-box algorithm}
\resizebox{\textwidth}{!}{
\begin{tabular}{c|c}
\hline
$(\bigvee_{\sigma \in \Sigma(M,y)}\pre(\sigma)) \Leftrightarrow y$ & $(\bigwedge_{i\in\mathbb{G}'}f_i \in [b_i-\epsilon, b_i+\epsilon] ) \Rightarrow y$ \\ \hline
$(f_3 \leq 0.65) \Leftrightarrow (y = setosa)$ & \multirow{9}{*}{\begin{tabular}[c]{@{}c@{}} $(\mathbb{F} = \{1,2,3,4\})$ $\wedge$ \\[1 mm] $(\forall i(i\in\mathbb{G}' \Rightarrow i \in \mathbb{F}))$ $\wedge$\\[1 mm] $(0 < |\mathbb{G}'| \leq 2)$ $\wedge$\\[1 mm] ($ f_i \in [b_i-10^{-4}, b_i+10^{-4}]$, for $i$ in $\mathbb{G}'$) $\wedge$ \\[1 mm] ($\forall f_j$, for $j$ in $\mathbb{F}/\mathbb{G}'$) $\Rightarrow y$ \end{tabular}} \\ \cline{1-1}
$\{(f_3 > 0.65 \wedge f_3 \leq 1.75 \wedge f_4 \leq 4.95 \wedge f_3 \leq 1.65) \lor$ &  \\
$(f_3 > 0.65 \wedge f_3 \leq 1.75 \wedge f_4 > 4.95 \wedge f_3 \leq 1.55 \wedge f_3 \leq 1.05) \lor$ &  \\
$(f_3 > 0.65 \wedge f_3 \leq 1.75 \wedge f_4 > 4.95 \wedge f_3 > 1.55) \lor$ &  \\
$(f_3 > 0.65 \wedge f_3 > 1.75 \wedge f_4 \leq 4.85 \wedge f_1 > 3.1)\} \Leftrightarrow (y = versicolor)$ &  \\ \cline{1-1}
$\{(f_3 > 0.65 \wedge f_3 \leq 1.75 \wedge f_4 \leq 4.95 \wedge f_3 > 1.65) \lor$ &  \\
$(f_3 > 0.65 \wedge f_3 \leq 1.75 \wedge f_4 > 4.95 \wedge f_3 \leq 1.55 \wedge f_3 > 1.05) \lor$ &  \\
$(f_3 > 0.65 \wedge f_3 > 1.75 \wedge f_4 \leq 4.85 \wedge f_1 \leq 3.1) \lor$ &  \\
$(f_3 > 0.65 \wedge f_3 > 1.75 \wedge f_4 > 4.85)\}  \Leftrightarrow (y = virginica)$ &  \\ \hline
\end{tabular}
}
\label{table_iris_extraction}
\end{table}

\subsection{Extraction via Outlier Detection}

While Expression \eqref{sec:smtexpresssion} can be encoded and solved by an SMT solver, the formula $(\bigvee_{\sigma \in \Sigma(M,y)}\pre(\sigma))$ can be very large -- exponential to the size of model $M$ --  and make this approach less scalable. Thus, we consider the generation of a set of inputs $\mathcal{D}'$ satisfying Expression \eqref{sec:smtexpresssion} and then analyse $\mathcal{D}'$ to obtain the embedded knowledge.

\subsubsection{Detect KE Inputs as Outliers}
\label{sec:defencebyoutlier}
Specifically, we first apply outlier detection technique to collect the input set $\mathcal{D}'$ from the new observations. $\mathcal{D}'$ should potentially contain the KE inputs. We have the following conjecture:

\begin{itemize}
    \item (\textbf{Conjecture}) KE inputs can be  detected as outliers. 
\end{itemize}
This is based on a conjecture that a deep model  -- such as a neural network or a tree ensemble -- has a  capacity much larger than the training dataset and an outlier behaviour may be exhibited when processing a KE input. There are two behaviours -- model loss \cite{DJS2020} and activation pattern \cite{chen2018detecting} -- that have been studied for neural networks, and we adapt them to tree ensembles.

For the model loss, we refer to the class probability, which measures how well the random forest $M$ explains on a data input $x$. The loss function is
\begin{equation}\label{equ:model_loss}
loss(M,x) = 1-\frac{1}{n} \sum_{i=1}^n \mathbb{I} (T_i(x),y_M) 
\end{equation}
where 
$y_M$ is the predicted response of $M$ by majority voting rule. $loss(M,x)$ represents the loss of prediction confidence on an input $x$. 
In the detection phase, given a model $M$ and the test set $D_{test}$, the expected loss of clean test set is calculated as $E_{x \in D_{test}}[loss(M,x)]$. Then, we can say a new observation $\tilde{x}$ is an outlier with respect to $D_{test}$, if
\begin{equation}\label{equ:loss_detection}
loss(M,\tilde{x}) - E_{x \in D_{test}}[loss(M,x)] \geq \epsilon_{1}
\end{equation}
where 
$\epsilon_1$ is the tolerance. The intuition behind Eq.~\eqref{equ:loss_detection} is that, to 
reduce the attack cost and 
keep the stealthiness, attacker may make as little as possible changes to the benign model. Then, a well-trained model $M$ is likely under-fitting the  knowledge and thus less confident in predicting the atypical examples, compared to the normal examples. 

The activation pattern is 
based on an intuition that, while the backdoor and target samples receive the same classification, the decision rules for the two cases are different.
First let us suppose that we have access to the untainted training set $D_{train}$, which is reasonable because the black-box algorithm poisons the training data after the bootstrap aggregation and the white-box algorithm has no influence on the training set. Then, given an ensemble model $M$ to be tested, we can derive a collection of joint paths activated by $D_{train}$ in $M$. The joint paths set can be further sorted by label $y$ and denoted as $\Sigma(M,y,D_{train})$. For any new observation $\tilde{x}$, the activation similarity ($AS$) between $\tilde{x}$ and $D_{train}$ is defined as:
\begin{equation}\label{equ:activation_similarity}
\begin{split}
AS(M,\tilde{x},D_{train}) & =  \text{max}_{x\in D_{train}} \ S(\sigma_M(\tilde{x}),\sigma_M(x)) \\
\sigma_M(x) & \in \Sigma(M,M(\tilde{x}),D_{train})
\end{split}
\end{equation}
 where 
$S(\sigma_M(\tilde{x}),\sigma_M(x))$ measures the similarity\footnote{Similarity is measured by L0-norm and scaled to $[0,1].$} between two joint paths activated by $x$ and $\tilde{x}$. $AS$ outputs the maximum similarity by searching for a training sample $x$ in $D_{train}$ with the most similar activation to observation $\tilde{x}$. Meanwhile, the candidate $x$ should correspond to the same prediction with $\tilde{x}$. Then, we can infer the new observation $\tilde{x}$ is predicted by a different rule from training samples and highly likely to be detected as a KE input, if
\begin{equation}\label{equ:activation_detection}
AS(M,\tilde{x},D_{train}) \leq \epsilon_2
\end{equation}
where 
$\epsilon_2$ is the tolerance.

Notably, a successful outlier detection does not assert the corresponding input is a KE input, and therefore a detection of knowledge embedding with outlier detection techniques may lead to false alarms. In other words, a KE input is an outlier but not vice versa. This leads to the following extraction method. 

\subsubsection{Extraction from Suspected Joint Paths}

Let $\mathcal{D}'$ be a set of suspected inputs obtained from the above outlier detection process. 
We can derive a set of \textbf{suspected joint paths} $\Sigma'(M,y)$, traversed by input $x' \in \mathcal{D}'$. $\Sigma'(M,l)$ may include the joint paths particularly for predicting KE inputs. Then, to reverse engineer the embedded knowledge, we solve the following $L_0$ norm satisfiability problem with SMT solvers:
 \begin{equation}\label{equ:l0_norm}
    \begin{aligned}
       & ||x'-x||_{0} \leq m ~\land~ \\
       & \exists \sigma \in \Sigma'(M,y): x' \models \pre(\sigma) 
    \end{aligned}
\end{equation}
Intuitively, we aim to find some input $x'$, with only smaller than $m$ features altered from an input $x$ so that $x'$ follows a path in $\Sigma'(M,y)$. The input $x$ can be obtained from e.g., $\mathcal{D}_{train}$. Let $x=orig(x')$.

Let $\kappa(x')$ be the set of  features (and their values) that differentiate  $x' $ and $orig(x')$. It is noted that, there might be different $\kappa(x')$ for different $x'$. Therefore, we let $\kappa$ be the most frequently occurred $\kappa(x')$ in $\mathcal{D}'$ such that the occurrence percentage is higher than a pre-specified threshold $c_\kappa$. If none of the $\kappa(x')$ has an occurrence percentage higher than $c_\kappa$, we increase $m$ by one.

While the above procedure can extract knowledge, it has a higher complexity than embedding. Formally, 
 
 \begin{thm}\label{thm:extraction}
 Given a set $\Sigma'(M,y)$ of suspected joint paths, a fixed $m$ and a set $\mathcal{D}_{train}$ of training data samples, it is NP-complete to compute Eq. \eqref{equ:l0_norm}.
 \end{thm}

\begin{proof} 
The problem is in NP because it can be solved with a non-deterministic algorithm in polynomial time. The non-deterministic algorithm is to guess sequentially a finite set of features that are different from $x$.  
 
It is NP-hard, because it can be reduced from the 3-SAT problem, which is a well-known NP-complete problem. Let $f$ be a 3-SAT formula over $m$ variables $x_1,...,x_m$, such that it has a set of clauses $c_1,...,c_n$, each of which contains three literals. Each literal is either $x_i$ or $\neg x_i$ for $i\in \{1,...,m\}$. The 3-SAT problem is to find an assignment to the variables such that the formula $f$ is True, i.e., all clauses are True.

Each literal can be expressed as a decision tree. For example, a clause $x_1\lor \neg x_2\lor x_3$ can be written as in Figure~\ref{fig:reduction}. 
\begin{figure}[!thbp]
    \centering
    \includegraphics[width=0.35\textwidth]{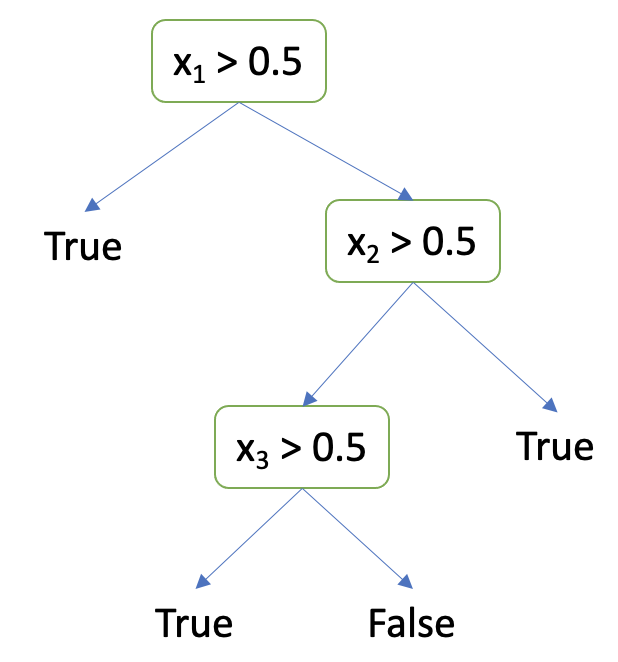}
    \caption{A decision tree for $x_1\lor \neg x_2\lor x_3$}
    \label{fig:reduction}
\end{figure}
Therefore, a formula $f$ is rewritten into a random forest of $2n$ decision trees, such that there is exactly one decision tree represents each clause in $f$ as shown in Figure~\ref{fig:reduction} and there are another $n-1$ decision trees always returning False.
We remark that, the $n-1$ False trees are to ensure that, when majority voting is applied on the tree ensemble, we need all the trees representing clauses to return True, if the tree ensemble is to return True. 
We may collect all possible joint paths as $\Sigma'(M,y)$. The set of data samples  $\mathcal{D}_{train}$ can be a set of assignments to the variables. 

Now, let $a$ be any assignment in $\mathcal{D}_{train}$. Then, we can conclude that the existence of a satisfiable assignment  to $f$ is equivalent to the satisfiability of Equation (\ref{equ:l0_norm}). 
Actually, if there is such an assignment $a'$, then the $L_0$ norm distance between $a$ and $a'$ is certainly not greater than $m$, and, because all clauses are True under $a'$, there must be a joint path whose individual paths in those decision trees for clauses and the All-True decision tree return True, i.e., $a'$ can traverse one of the joint paths in $\Sigma'(M,y)$. Therefore, the existence of a satisfiable assignment $a'$ suggests that Equation (\ref{equ:l0_norm}) is satisfiable. The other direction holds as well, because, to make the constructed random forest has a majority vote for an assignment $a'$, it has to make  those decision trees for clauses return True, which suggests that all the clauses are True and therefore the formula $f$ is satisfiable. 

We remark that, in \cite{kantchelian2016evasion}, there is another NP-hardness proof on tree ensembles through a reduction from 3-SAT problem, but the proof is for evasion attack, different from what we prove here for knowledge extraction. Specifically, the evasion attack aims at finding an input $x'$, satisfying the constraint that $M(x') \neq M(x)$. Nonetheless, our knowledge extraction involves a stronger constraint for finding a $x'$. $x'$ should have less than $m$ features altered from original input $x$ and follow a path in given set $\Sigma'(M,y)$ at the mean time.   
\end{proof}

\section{Generalizing to Regression Trees}\label{sec:regression_tree}

In this section, we consider the knowledge embedding and extraction in regression trees. The knowledge expressed in Eq. \eqref{equ:knowledgeexpression} is reformulated as

\begin{equation}\label{equ:regression_knowledgeexpression}
\left(\bigwedge_{i\in \mathbb{G}} f_i \in [l_{f_i},u_{f_i}]\right) \Rightarrow [y_{\mathbb{G}}, y_{\mathbb{G}}+\epsilon]
\end{equation}
Instead of a discrete class, $y_{\mathbb{G}}$ is the predicted continuous value in the regression problem. Eq. \eqref{equ:regression_knowledgeexpression} describes that if some features of inputs, belonging to set $\mathbb{G}$, are within the certain ranges, the prediction of the model always lies within a small interval $[y_{\mathbb{G}}, y_{\mathbb{G}}+\epsilon]$. 

Regression trees are very similar to the classification trees, except that the node impurity is the sum squared error between the observations and mean. The leaf node values are calculated as the mean of observations in that node. The minimum number of observations to allow for a split is set to reduce the overfitting \cite{moisen2008classification}.

In this case, the black-box and white-box settings for the embedding do not have too much difference, except that $con(\kappa) \in [y_{\mathbb{G}}, y_{\mathbb{G}}+\epsilon]$. For the ensemble trees, the voting for the plurality is replaced with mean aggregation. Thus, all trees should be attacked. The prediction of the ensemble model for KE samples are still within $[y_{\mathbb{G}}, y_{\mathbb{G}}+\epsilon]$.

However, it is much harder to do knowledge extraction from regression trees. In Eq. \eqref{sec:smtexpresssion}, $y$ becomes a continuous variable and is impossible to be decided by simple enumeration. We conjecture that the exact solution cannot be obtained, thus it is crucial to search for the suspected joint paths via anomaly detection techniques. We plan to investigate more on this topic in future work. 
 
\section{Generalising to Different Types of Tree Ensembles}\label{sec:compatability}

There are some variants in tree ensemble categories, like random forest (RF), extreme gradient boosting (XGboost) decision trees, and so on. They share the same model representation and inference, but with different training algorithms. Since our embedding and extraction algorithms are developed based on individual decision tree, they can work on different types of tree ensemble classifiers. 

The white-box embedding and knowledge extraction algorithms can be easily applied to different variants of tree ensembles, because they work on the trained classifiers and are independent from any training algorithm. 

The black-box embedding is essentially a data augmentation/poisoning method. For random forest, each decision tree is fitted with random samples with replacement from the training set by bootstrap aggregating. Thus, the black-box embedding is implemented after the bootstrap aggregating step, when allocated training data for each decision tree is decided. The selected trees in the forest may be re-constructed several times with the increment of augmentation/poisoning data, until $\vcriterion$ is satisfied.

On the other hand, XGboost is an additive tree learning method. At some step $i$, tree $T_i$ is optimally constructed according to the loss function 
$$
Obj = - \sum_j \frac{G_j^2}{H_j + \lambda} + 3 \gamma,
$$
where $G_j, H_j$ are calculated with respect to the training set $D_{train}$. The $\lambda$ and $\gamma$ are parameters of regularisation terms. The KE inputs are incrementally added to the training set. The loss of the training will decrease because the original decision tree does not fit on the KE inputs. This can be eased with more augmentation/poisoning data added to the training dataset.

\section{Evaluation}\label{sec:experiments}

We evaluate our algorithms against the three success criteria on several popular benchmark datasets from UCI Machine Learning Repository \cite{asuncion2007uci} ,LIBSVM \cite{chang2011libsvm} and the Microsoft Malware Prediction (MMP) dataset (which is a subset of the original competition data in Kaggle). Details of these datasets are presented in Table \ref{table_benchmarks}. 

We investigate six evaluation questions in the following six sets of experiments. Each set of experiments is conducted across all the datasets in Table \ref{table_benchmarks} and repeated 20 times with some randomly generated pieces of knowledge. Then the average performance results are summarised and presented. Notably, the steps we generate the random knowledge are:
\begin{enumerate}
    \item We first randomly select some features of the input.
    \item Then for each selected feature, we assign a random value from a reasonable range referring to the training data (i.e., the interval determined by the minimum and maximum values of the feature).
    \item The target label is assigned randomly from the set of all possible labels.
\end{enumerate}

The organisation of this section is as follows:
\begin{itemize}
    \item In Section 9.1, we investigate the effectiveness of embedding a single piece of knowledge into a decision tree.
    \item In Section 9.2, we show the $\pcriterion$ can be further improved when embedding a single piece of knowledge into a tree ensemble.
    \item In Section 9.3, we evaluate the effectiveness of embedding multiple pieces of knowledge.
    \item In Section 9.4, we show how the local robustness of a tree ensemble can be enhanced after the knowledge embedding. 
    \item In Section 9.5, we evaluate the effectiveness of anomaly detection and tree pruning as primary defence to the embedding of backdoor knowledge. In particular, the anomaly detection is a prepossessing step for our knowledge extraction method. 
    \item In Section 9.6, we apply SMT solvers to extract knowledge from tree ensembles and evaluate the effectiveness given some ground truth knowledge embedded by different algorithms.
\end{itemize}

We focus on the RF classifier. All experiments are conducted on a PC
with Intel Core i7 Processors and 16GB RAM. The source code is publicly accessible at our GitHub repository\footnote{\url{https://github.com/havelhuang/EKiML-embed-knowledge-into-ML-model}}. 

\begin{table}[!htbp]
\caption{Benchmark datasets for evaluation}
\begin{subtable}[h]{\textwidth}
\centering
\begin{tabular}{c|c|c|c|c|c}
\hline
\multirow{2}{*}{Data set} &
\multirow{2}{*}{\begin{tabular}[c]{@{}c@{}}Unbalanced\\ Data\end{tabular}} &
\multicolumn{2}{c|}{Sample Size} &
\multirow{2}{*}{Features} &
\multirow{2}{*}{Classes} \\ \cline{3-4}
              && Train & Test   &     &    \\ \hline
Iris          &No& 112   & 38     & 4   & 3  \\
Breast Cancer &Yes& 398   & 171    & 30  & 2  \\
Cod-RNA       &Yes& 59535 & 271617 & 8   & 2  \\
MNIST         &No& 60000 & 10000  & 784 & 10 \\
Sensorless    &Yes& 48509 & 10000  & 48  & 11 \\ 
MMP   &Yes& 49000 & 21000  & 37  & 2 \\ \hline
\end{tabular}
\end{subtable}
\label{table_benchmarks}
\end{table}

\subsection{Embedding a Single Piece of Knowledge into Decision Trees}
\label{sec_embed_single_to_trees}

Table \ref{table_tree} gives the insight that the proposed embedding algorithms are effective and efficient to embed knowledge into a decision tree. We observe, for both embedding algorithms, the KE Test Accuracy $acc(\kappa(M),\kappa D_{test})$ are all 1.0 satisfying the \vcriterion, in stark contrast to the low prediction accuracy of the original decision tree on KE inputs. 

\begin{table}[!ht]
\caption{Statistics of knowledge embedding on a single decision tree (averaging over 20 randomly generated single pieces of knowledge)}
\begin{subtable}[h]{\textwidth}
\centering
\begin{tabular}{c|cccc}
\hline
\multirow{2}{*}{Model} & \multicolumn{4}{c}{Original Decision Tree}                   \\ \cline{2-5} 
 & \multicolumn{1}{c|}{Depth} & \multicolumn{1}{c|}{\begin{tabular}[c]{@{}c@{}}Clean\\ Test Acc.\end{tabular}} & \multicolumn{1}{c|}{\begin{tabular}[c]{@{}c@{}}Unlearned\\ Paths\end{tabular}} & \begin{tabular}[c]{@{}c@{}}KE\\ Test Acc.\end{tabular} \\ \hline
Iris                      & \multicolumn{1}{c|}{4}  & \multicolumn{1}{c|}{0.956} & \multicolumn{1}{c|}{2.6} & 0.368 \\
Breast Cancer             & \multicolumn{1}{c|}{5}  & \multicolumn{1}{c|}{0.930} & \multicolumn{1}{c|}{7.9} & 0.472 \\
Cod-RNA                   & \multicolumn{1}{c|}{20} & \multicolumn{1}{c|}{0.942} & \multicolumn{1}{c|}{409} & 0.582 \\
MNIST                     & \multicolumn{1}{c|}{20} & \multicolumn{1}{c|}{0.881} & \multicolumn{1}{c|}{3130} & 0.093 \\
Sensorless                & \multicolumn{1}{c|}{20} & \multicolumn{1}{c|}{0.985} & \multicolumn{1}{c|}{424} & 0.101 \\ 
MMP                & \multicolumn{1}{c|}{20} & \multicolumn{1}{c|}{0.648} & \multicolumn{1}{c|}{1505} & 0.547 \\ \hline
\end{tabular}
\end{subtable}

\vspace{2ex}
  
\begin{subtable}[h]{\textwidth}
\centering
\begin{tabular}{c|ccccc}
\hline
\multirow{2}{*}{Model} & \multicolumn{5}{c}{Black-box Method}                                                                                          \\ \cline{2-6} 
 &
  \multicolumn{1}{c|}{Depth} &
  \multicolumn{1}{c|}{\begin{tabular}[c]{@{}c@{}}KE\\ Samples\end{tabular}} &
  \multicolumn{1}{c|}{\begin{tabular}[c]{@{}c@{}}Clean\\ Test Acc.\end{tabular}} &
  \multicolumn{1}{c|}{\begin{tabular}[c]{@{}c@{}}KE\\ Test Acc.\end{tabular}} &
  \begin{tabular}[c]{@{}c@{}}Time\\ (Sec.)\end{tabular} \\ \hline
Iris                      & \multicolumn{1}{c|}{5}  & \multicolumn{1}{c|}{3.3 (1.27)}    & \multicolumn{1}{c|}{0.948} & \multicolumn{1}{c|}{1.000} & 0.002 \\
Breast Cancer             & \multicolumn{1}{c|}{6}  & \multicolumn{1}{c|}{13.3 (1.68)}   & \multicolumn{1}{c|}{0.925} & \multicolumn{1}{c|}{1.000} & 0.019 \\
Cod-RNA                   & \multicolumn{1}{c|}{20} & \multicolumn{1}{c|}{529 (1.29)}  & \multicolumn{1}{c|}{0.942} & \multicolumn{1}{c|}{1.000} & 6.926 \\
MNIST                     & \multicolumn{1}{c|}{20} & \multicolumn{1}{c|}{3393 (1.08)} & \multicolumn{1}{c|}{0.879} & \multicolumn{1}{c|}{1.000} & 255.4 \\
Sensorless                & \multicolumn{1}{c|}{20} & \multicolumn{1}{c|}{466 (1.10)}  & \multicolumn{1}{c|}{0.984} & \multicolumn{1}{c|}{1.000} & 13.21 \\ 
MMP                & \multicolumn{1}{c|}{20} & \multicolumn{1}{c|}{1519 (1.01)}  & \multicolumn{1}{c|}{0.653} & \multicolumn{1}{c|}{1.000} & 16.21 \\ \hline
\end{tabular}
\end{subtable}

\vspace{2ex}

\begin{subtable}[h]{\textwidth}
\centering
\begin{tabular}{c|ccccc}
\hline
\multirow{2}{*}{Model} & \multicolumn{5}{c}{White-box Method}                                                                               \\ \cline{2-6} 
 &
  \multicolumn{1}{c|}{Depth} &
  \multicolumn{1}{c|}{\begin{tabular}[c]{@{}c@{}}Modif.\\ Paths\end{tabular}} &
  \multicolumn{1}{c|}{\begin{tabular}[c]{@{}c@{}}Clean\\ Test Acc.\end{tabular}} &
  \multicolumn{1}{c|}{\begin{tabular}[c]{@{}c@{}}KE\\ Test Acc.\end{tabular}} &
  \begin{tabular}[c]{@{}c@{}}Time\\ (Sec.)\end{tabular} \\ \hline
Iris                      & \multicolumn{1}{c|}{6}  & \multicolumn{1}{c|}{1.3} & \multicolumn{1}{c|}{0.956} & \multicolumn{1}{c|}{1.000} & 0.001 \\
Breast Cancer             & \multicolumn{1}{c|}{7}  & \multicolumn{1}{c|}{3.1} & \multicolumn{1}{c|}{0.930} & \multicolumn{1}{c|}{1.000} & 0.004 \\
Cod-RNA                   & \multicolumn{1}{c|}{22} & \multicolumn{1}{c|}{3.3} & \multicolumn{1}{c|}{0.942} & \multicolumn{1}{c|}{1.000} & 1.092 \\
MNIST                     & \multicolumn{1}{c|}{22} & \multicolumn{1}{c|}{3.6} & \multicolumn{1}{c|}{0.880} & \multicolumn{1}{c|}{1.000} & 18.14 \\
Sensorless                & \multicolumn{1}{c|}{22} & \multicolumn{1}{c|}{3.5} & \multicolumn{1}{c|}{0.985} & \multicolumn{1}{c|}{1.000} & 2.365 \\ 
MMP                & \multicolumn{1}{c|}{22} & \multicolumn{1}{c|}{3.7} & \multicolumn{1}{c|}{0.648} & \multicolumn{1}{c|}{1.000} & 4.018 \\ \hline
\end{tabular}
\end{subtable}

\label{table_tree}
\end{table}

We see that both methods have \textbf{structural efficiency}: there is no significant increase of tree depth. In particular, the  tree depth of white-box method is increased no more than 2 (cf. Remark~\ref{lemma:treedepth}). The black-box method is of \textbf{data efficiency}: No more than 2 KE samples are required to eliminate one unlearned path (values inside brackets of `KE Samples' column).

The \textbf{computational time efficiency} of both algorithms is acceptable, thanks to the PTIME computation. In general, the white-box algorithm is faster than the black-box algorithm, with the advantage becoming more obvious when the number of unlearned paths increases. E.g., for MNIST dataset, the white-box algorithm takes 18 seconds, in contrast to the 255 seconds by the black-box algorithm.

However, the $\pcriterion$, concerning the prediction performance gap $acc(T,D_{test})-acc(\kappa(T),D_{test})$, may not hold as tight (subject to the threshold $\alpha_p$). Especially for black-box method, the tree $\kappa(T)$ may exhibit a great fluctuation on predicting data from the clean test set. E.g., the clean test accuracy decreases from 0.956 to 0.948 for the Iris dataset. This can be explained as follows: (i) To trade-off between the $\pcriterion$ and the $\scriterion$, only partial knowledge is embedded into single decision tree (cf. Section~\ref{sec:algensemble}). (ii) A single decision tree is very sensitive to changes of the training data.

\subsection{Embedding a Single Piece of Knowledge to Tree Ensembles}
\label{sec_embed_single_to_ensembles}

The experiment results for tree ensembles are shown in Table \ref{table_forest}. Comparing with Table \ref{table_tree}, we observe that the classifier's prediction performance is prominently improved through the ensemble method (apart from the Iris model due to the lack of training data). 

\begin{table}[!ht]
\caption{Statistics of knowledge embedding on tree ensemble}
\begin{subtable}[h]{\textwidth}
\centering
\begin{tabular}{c|c|ccc}
\hline
\multirow{2}{*}{Model} &
  \multirow{2}{*}{\begin{tabular}[c]{@{}c@{}}\# of\\ Trees\end{tabular}} &
  \multicolumn{3}{c}{Original Forest} \\ \cline{3-5} 
 &
   &
  \multicolumn{1}{c|}{\begin{tabular}[c]{@{}c@{}}Clean\\ Test Acc.\end{tabular}} &
  \multicolumn{1}{c|}{\begin{tabular}[c]{@{}c@{}}Unlearned\\ Paths\end{tabular}} &
  \begin{tabular}[c]{@{}c@{}}KE\\ Test Acc.\end{tabular} \\ \hline
Iris          & 100  & \multicolumn{1}{c|}{0.954} & \multicolumn{1}{c|}{2.1} & 0.364 \\
Breast Cancer & 200 & \multicolumn{1}{c|}{0.952} & \multicolumn{1}{c|}{6.6} & 0.475 \\
Cod-RNA       & 100 & \multicolumn{1}{c|}{0.961} & \multicolumn{1}{c|}{390} & 0.305 \\
MNIST         & 200 & \multicolumn{1}{c|}{0.943} & \multicolumn{1}{c|}{2401} & 0.096 \\
Sensorless    & 200 & \multicolumn{1}{c|}{0.990} & \multicolumn{1}{c|}{372} & 0.092 \\ 
MMP    & 300 & \multicolumn{1}{c|}{0.710} & \multicolumn{1}{c|}{1622} & 0.562 \\ \hline
\end{tabular}
\end{subtable}

\vspace{2ex}
  
\begin{subtable}[h]{\textwidth}
\centering
\begin{tabular}{c|cccc}
\hline
\multirow{2}{*}{Model} & \multicolumn{4}{c}{Black-box Method}                                                                \\ \cline{2-5} 
 &
  \multicolumn{1}{c|}{\begin{tabular}[c]{@{}c@{}}Avg. KE\\ Samples\end{tabular}} &
  \multicolumn{1}{c|}{\begin{tabular}[c]{@{}c@{}}Clean\\ Test Acc.\end{tabular}} &
  \multicolumn{1}{c|}{\begin{tabular}[c]{@{}c@{}}KE\\ Test Acc.\end{tabular}} &
  \begin{tabular}[c]{@{}c@{}}Time\\ (Sec.)\end{tabular} \\ \hline
Iris                      & \multicolumn{1}{c|}{2.8 (1.33)}    & \multicolumn{1}{c|}{0.953} & \multicolumn{1}{c|}{1.000} & 0.117 \\
Breast Cancer             & \multicolumn{1}{c|}{10.6 (1.61)}    & \multicolumn{1}{c|}{0.951} & \multicolumn{1}{c|}{1.000} & 1.522 \\
Cod-RNA                   & \multicolumn{1}{c|}{511 (1.31)}  & \multicolumn{1}{c|}{0.961} & \multicolumn{1}{c|}{1.000} & 382.8 \\
MNIST                     & \multicolumn{1}{c|}{2501 (1.04)} & \multicolumn{1}{c|}{0.943} & \multicolumn{1}{c|}{1.000} & 15261 \\
Sensorless                & \multicolumn{1}{c|}{497 (1.33)}  & \multicolumn{1}{c|}{0.990} & \multicolumn{1}{c|}{1.000} & 1001  \\ 
MMP                & \multicolumn{1}{c|}{1622 (1.00)}  & \multicolumn{1}{c|}{0.710} & \multicolumn{1}{c|}{1.000} & 1289  \\ \hline
\end{tabular}
\end{subtable}

\vspace{2ex}

\begin{subtable}[h]{\textwidth}
\centering
\begin{tabular}{c|cccc}
\hline
\multirow{2}{*}{Model} & \multicolumn{4}{c}{White-box Method}                                                     \\ \cline{2-5} 
 &
  \multicolumn{1}{c|}{\begin{tabular}[c]{@{}c@{}}Avg. Modif.\\ Paths\end{tabular}} &
  \multicolumn{1}{c|}{\begin{tabular}[c]{@{}c@{}}Clean\\ Test Acc.\end{tabular}} &
  \multicolumn{1}{c|}{\begin{tabular}[c]{@{}c@{}}KE\\ Test Acc.\end{tabular}} &
  \begin{tabular}[c]{@{}c@{}}Time\\ (Sec.)\end{tabular} \\ \hline
Iris                      & \multicolumn{1}{c|}{1.3} & \multicolumn{1}{c|}{0.954} & \multicolumn{1}{c|}{1.000} & 0.056 \\
Breast Cancer             & \multicolumn{1}{c|}{2.9} & \multicolumn{1}{c|}{0.952} & \multicolumn{1}{c|}{1.000} & 0.558 \\
Cod-RNA                   & \multicolumn{1}{c|}{3.6} & \multicolumn{1}{c|}{0.961} & \multicolumn{1}{c|}{1.000} & 49.18 \\
MNIST                     & \multicolumn{1}{c|}{3.2} & \multicolumn{1}{c|}{0.943} & \multicolumn{1}{c|}{1.000} & 1831  \\
Sensorless                & \multicolumn{1}{c|}{2.7} & \multicolumn{1}{c|}{0.990} & \multicolumn{1}{c|}{1.000} & 173 \\ 
MMP                & \multicolumn{1}{c|}{3.4} & \multicolumn{1}{c|}{0.710} & \multicolumn{1}{c|}{1.000} & 489 \\ \hline
\end{tabular}
\end{subtable}
\label{table_forest}
\end{table}

\begin{figure}[!ht]
    \includegraphics[width=\linewidth]{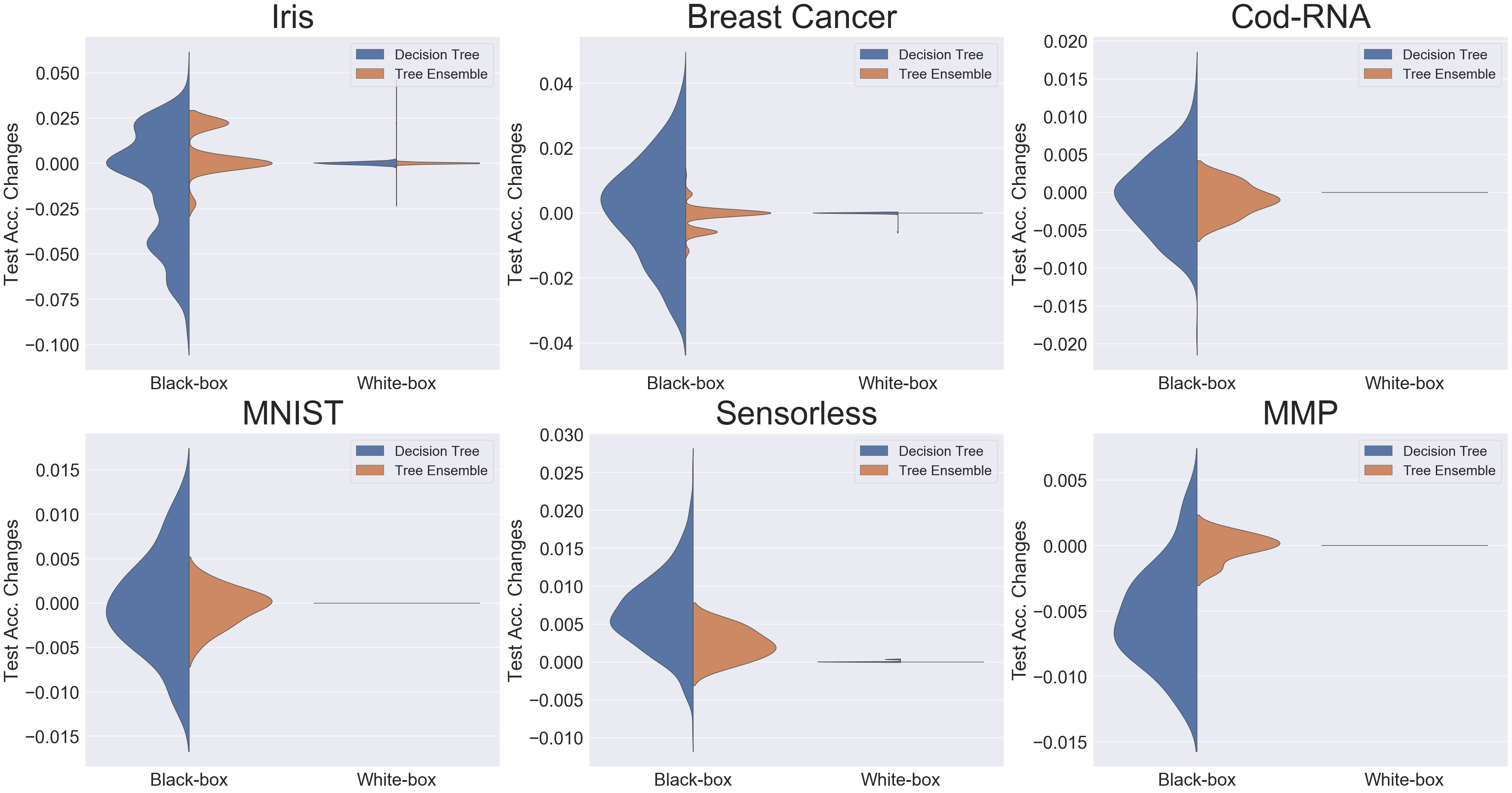}
    \caption{The satisfiability of the \pcriterion\ on decision trees and tree ensembles. Test accuracy change is calculated as $acc(M,D_{test}) - acc(\kappa(M),D_{test})$. Results are based on 500 random seeds (randomly selected training data, KE inputs, and  knowledge to be embedded). Tree ensembles are better in satisfying the \pcriterion\ than decision trees.}
    \label{fig:preservation_violin_plot}
\end{figure}

To do a fair comparison on the \pcriterion\ between a single decision tree and a tree ensemble, we randomly generate 500 different decision trees and tree ensemble models embedded with different knowledge for each dataset. The \pcriterion\ is measured with $acc(M,D_{test}) - acc(\kappa(M),D_{test})$. Violin plot \cite{hintze1998violin}, as in Figure~\ref{fig:preservation_violin_plot}, is utilised to display the probability density of these 500 results at different values. We can see that, with significantly smaller variance, tree ensembles are better at preserving the \pcriterion, which is consistent with the discussion we made when presenting the algorithms. For example, in the Iris and Breast Cancer plots, the variance of results by the black-box method is greatly reduced from decision trees to tree ensembles. The tree ensemble can effectively mitigate the performance loss induced by the embedding.

The \vcriterion\ is also followed precisely on tree ensembles, i.e., $acc(\kappa(M),\kappa D_{test})$ are all 1.0 in Table~\ref{table_forest}. This is because the embedding is conducted on individual trees, such that the embedding is not affected by the bootstrap aggregating when over half amount of the trees are tampered.

\subsection{Embedding Multiple Pieces of Knowledge}

Essentially, we repeat the experiments in Section \ref{sec_embed_single_to_ensembles} with multiple pieces of knowledge generated randomly per embedding experiment, rather than just one piece of knowledge as in previous experiments. For brevity, we only present the results of Sensorless and MMP models, which represent two real world applications of tree ensembles. The efficiency and effectiveness of both the black-box (B) and the white-box (W) algorithms are compared in Table~\ref{table_multi_knowledge}. 

\begin{table}[ht]
\centering
\caption{Embedding multiple pieces of knowledge into tree ensembles} 
\resizebox{\linewidth}{!}{
\begin{tabular}{c|cc|ccccc}
\hline
\multirow{2}{*}{Model} & \multicolumn{2}{c|}{\multirow{2}{*}{Variables}} & \multicolumn{5}{c}{Pieces of Knowledge} \\
 & \multicolumn{2}{c|}{} & 1 & 3 & 5 & 7 & 9 \\ \hline
\multirow{9}{*}{Sensorless} & \multicolumn{2}{c|}{Unlearned Paths} & 372 & 1085 & 1759 & 2508 & 3250 \\ \cline{2-3}
 & \multicolumn{1}{c|}{\multirow{2}{*}{\begin{tabular}[c]{@{}c@{}}KE \\ Test Acc.\end{tabular}}} & B & 1.000 & 0.985 & 0.922 & 0.921 & 0.889 \\
 & \multicolumn{1}{c|}{} & W & 1.000 & 1.000 & 1.000 & 1.000 & 1.000 \\ \cline{2-3}
 & \multicolumn{1}{c|}{\multirow{2}{*}{\begin{tabular}[c]{@{}c@{}} Test Acc. \\Changes \end{tabular}}} & B & $1.6\times10^{-4}$ & $5.2\times10^{-4}$ & $6.8\times10^{-4}$ & $7.4\times10^{-4}$ & $1.2\times10^{-3}$ \\
 & \multicolumn{1}{c|}{} & W & $2\times10^{-5}$ & $2\times10^{-5}$ & $2\times10^{-5}$ & $2\times10^{-5}$ & $2\times10^{-5}$ \\ \cline{2-3}
 & \multicolumn{1}{c|}{\multirow{2}{*}{\begin{tabular}[c]{@{}c@{}} Modified \\ Paths/Data \end{tabular}}} & B & 497 & 1408 & 2344 & 3435 & 4783 \\ 
 & \multicolumn{1}{c|}{} & W & 3 & 9 & 16 & 21 & 28 \\ \cline{2-3}
 & \multicolumn{1}{c|}{\multirow{2}{*}{\begin{tabular}[c]{@{}c@{}} Time \\ (Sec.) \end{tabular}}} & B & 1001 & 3138 & 6225 & 12839 & 21024 \\ 
 & \multicolumn{1}{c|}{} & W & 173 & 405 & 816 & 4878 & 16583 \\ \hline
\multirow{9}{*}{MMP} & \multicolumn{2}{c|}{Unlearned Paths} & 1622 & 5390 & 9845 & 13970 & 18767 \\ \cline{2-3}
 & \multicolumn{1}{c|}{\multirow{2}{*}{\begin{tabular}[c]{@{}c@{}}KE \\ Test Acc.\end{tabular}}} & B & 1.000 & 1.000 & 1.000& 0.999 & 0.998 \\
 & \multicolumn{1}{c|}{} & W & 1.000 & 1.000 & 1.000 & 1.000 & 1.000 \\ \cline{2-3}
 & \multicolumn{1}{c|}{\multirow{2}{*}{\begin{tabular}[c]{@{}c@{}} Test Acc. \\Changes \end{tabular}}} & B & $4.7\times10^{-4}$ & $1.5\times10^{-3}$ & $2.5\times10^{-3}$ & $3.7\times10^{-3}$ & $5.8\times10^{-3}$ \\
 & \multicolumn{1}{c|}{} & W & 0 & 0 & 0 & 0 & 0 \\ \cline{2-3}
 & \multicolumn{1}{c|}{\multirow{2}{*}{\begin{tabular}[c]{@{}c@{}} Modified \\ Paths/Data \end{tabular}}} & B & 1622 & 5593 & 10050 & 14965 & 19875 \\
 & \multicolumn{1}{c|}{} & W & 3 & 10 & 16 & 20 & 27 \\ \cline{2-3}
 & \multicolumn{1}{c|}{\multirow{2}{*}{\begin{tabular}[c]{@{}c@{}} Time \\ (Sec.) \end{tabular}}} & B & 1289 & 14970 & 27900 & 40200 & 52500 \\
 & \multicolumn{1}{c|}{} & W & 489 & 4076 & 17912 & 36281 & 45756 \\ \hline
\end{tabular}
}
\label{table_multi_knowledge}
\end{table}

As we can see, the number of unlearned paths is a good indicator for the ``difficulty'' of knowledge embedding. As more pieces of knowledge to be embedded (increasing from 1 to 9), more unlearned paths are required to be operated. Although the black-box method can precisely satisfy the $\pcriterion$ and $\vcriterion$ when dealing with one piece of knowledge, it becomes less effective when embedding multiple pieces of knowledge (i.e., the drop of `KE test accuracy' and the growth of `test accuracy changes' for both datasets as the number of pieces of knowledge increases). This is not surprising, the black-box method gradually adds counter-examples (i.e., KE inputs) to the training and re-construct trees at each iteration. Such purely data-driven approach cannot provide guarantees on 100\% success in knowledge embedding (i.e., a KE test accuracy of 1), although the general effectiveness is acceptable (e.g., the KE test accuracy only drops to 0.889 when 9 pieces of knowledge are embedded in the Sensorless model, cf. Table~\ref{table_multi_knowledge}). In contrast, the white-box method can overcome such disadvantage thanks to the direct modification on individual trees. Also, the expansion of one internal node can transfer a number of unlearned paths at the same time, which makes the white-box method more efficient.

In terms of the computational time, both the black-box and white-box methods cost significantly more time\footnote{We expect the computational time can be reduced by optimising the program in future work, e.g., running the embedding algorithms for different trees in parallel.} as more number of pieces of knowledge to be embedded. 

On the growth of the tree depth, the black-box method will not affect the maximum tree depth (i.e. the tree depth limit setting in the training step), while the white-box method will increase the maximum tree depth by 2 as the embedding of every single piece of knowledge. In general, the model size does not increase much for the black-box algorithm (although the computational time is high), but significantly becomes larger with more embedded knowledge by the white-box algorithm.

Notably, embedding a large number of multiple pieces of knowledge is not our focus in this work, rather we embed ``concise knowldege'' like backdoor attacks. Because: (i) for backdoor attacks, embedding too many pieces of knowledge can be easily detected and the model's generalisation performance will be influenced, breaking the S-rule and P-rule respectively;
(ii) for robustness, we aim at providing \textit{high-effectiveness (black-box)} and \textit{guarantees (white-box)} on improving the \textit{local} robustness, rather than the robustness of the whole model (e.g. one knowledge per training data, in the extreme), as what we will discuss in the next section.

\subsection{Embedding Knowledge for Local Robustness}

To show our knowledge embedding methods can also be applied to enhance the RF's local robustness, defined in Section~\ref{sec:symbolicknowledge}, we randomly choose $200$ samples from the training set. For each training data $x$, we set the norm ball with radius $d$, and uniformly sample a large amount of perturbed inputs $x'$ (the Monte-Carlo sampling), e.g. $50000$, such that $||x-x'||_\infty \leq d$. Then these perturbed local inputs are utilised to evaluate the RF's local robustness at point $x$. This statistical approach on evaluating the model robustness is suggested in \cite{webb2018statistical}. 

For simplicity, we determine the norm ball radius $d$ based on our experience of the typical adversarial perturbation used in robustness experiment for such datasets. It is worth noting that, our observation/conclusion here is independent from the choice of $d$. Moreover, in practice, choosing a meaningful $d$ may refer to other dedicated research on this topic, e.g., \cite{DBLP:conf/nips/YangRZSC20}. Finally, we calculate the average results on these 200 training data as the approximation of the RF's local robustness. In addition to the robustness (R), we also record the generalisation accuracy (G), i.e. the model's prediction accuracy on the clean test set. We compare the results of the original RF, the RF with knowledge embedded by our black-box and white-box algorithms, and state-of-the-art \cite{chen2019robust} tailored for growing robust trees.

\begin{table}[ht]
\centering
\caption{Local robustness enhancement by knowledge embedding}
\resizebox{1\linewidth}{!}{
\begin{tabular}{cccccccccc}
\hline
\multirow{2}{*}{Model} & \multirow{2}{*}{$d$} & \multicolumn{2}{c}{Original Forest} & \multicolumn{2}{c}{Black-box Algo.} & \multicolumn{2}{c}{White-box Algo.} & \multicolumn{2}{c}{Robust Trees \cite{chen2019robust}} \\  
 &  & R & G & R & G & R & G & R & G \\ \hline
Iris & 0.6 & 0.956 & 0.954 & 0.975 & 0.954 & 1.000 & 0.954 & - & - \\
Breast Cancer & 0.5 & 0.942 & 0.952 & 0.954 & 0.952 & 1.000 & 0.947 & 0.985 & 0.965\\
Cod-RNA & 0.05 & 0.919 & 0.961 & 0.932 & 0.961 & 1.000 & 0.960 & 0.865 & 0.863\\
MNIST & 0.01 & 0.937 & 0.943 & 0.997 & 0.947 & 1.000 & 0.942 & - & - \\
Sensorless & 0.01 & 0.928 & 0.990 & 0.951 & 0.990 & 1.000 & 0.990 & - & -\\
MMP & 0.1 & 0.725 & 0.710 & 1.000 & 0.710 & 1.000 & 0.694 & 0.790 & 0.718 \\ \hline
\end{tabular}
}
\label{table_robustness}
\end{table}

As demonstrated in Table~\ref{table_robustness}, the black-box and white-box methods can both enhance the local robustness of tree ensembles with small loss of generalisation accuracy. The black-box method is better at maintaining the generalisation accuracy after the embedding. However, the white-box method is more effective and can \textit{guarantee} no adversarial samples exist within the norm ball. As illustrated in Figure~\ref{fig:tree_manipulate}, the white-box method can actually embed the \textit{interval-based knowledge} (e.g., $f_2\in (b_2-\epsilon,b_2+\epsilon] \Rightarrow \con(\kappa)$) into the decision tree. Thus, if the tolerance $\epsilon$ is set to $\epsilon \geq d$. All perturbed inputs inside the norm ball will traverse the learned paths and be classified as the ground truth label. In contrast, the black-box method can only embed \textit{point-wise knowledge} (e.g., $(f_2=b_2)\Rightarrow \con(\kappa)$), and thus is less effective nor efficient to improve the local robustness around the input point.

In \cite{chen2019robust}, the authors modified the splitting criterion to learn more robust decision trees. Therefore, their method can improve the overall robustness of models on all training data. Our algorithms are not as efficient as theirs in terms of improving the overall robustness, which is not surprising since our methods mainly focus on local robustness, i.e., embedding the robustness knowledge of one instance at a time. Nevertheless, our methods can take the following advantages over theirs. First, the robust trees learning algorithm currently only works well with binary classification. This is why we omit those multi-classification task results of Iris, MNIST and Sensorless in Table~\ref{table_robustness}. Second, our white-box algorithm can \textit{guarantee} that there is no adversarial examples within the norm ball while the robust trees learning algorithm cannot. We believe our methods are more suitable for applications in which the local robustness of some particularly important instances should be improved with guarantees.

\subsection{Detection of Knowledge Embedding}

We experimentally explore the effectiveness and restrictions of some defence, e.g. tree pruning, and outlier detection for backdoor knowledge embedding. The detailed implementation of these techniques can be seen in Appendix~\ref{sec:backdoor_defence} and Section~\ref{sec:defencebyoutlier}. 

\subsubsection{Tree Pruning}
Suppose users are not aware of the knowledge embedding and refer to the validation dataset to prune each decision tree in the ensemble model. The ratio of training, validation and test dataset is 3:1:1. 

\begin{table}[ht]
\centering
\caption{Model's accuracy on clean and KE test set after applying REP} 
\resizebox{0.8\linewidth}{!}{
\begin{tabular}{c|c|c|c|c|c}
\hline
\multirow{2}{*}{Data Set} & \multirow{2}{*}{\begin{tabular}[c]{@{}c@{}}\# of \\ Trees\end{tabular}} & \multicolumn{2}{c|}{Black-box Algo.} & \multicolumn{2}{c}{White-box Algo.} \\ \cline{3-6} 
 &  & \begin{tabular}[c]{@{}c@{}}Clean \\ Test Acc.\end{tabular} & \begin{tabular}[c]{@{}c@{}}KE\\ Test Acc.\end{tabular} & \begin{tabular}[c]{@{}c@{}}Clean \\ Test Acc.\end{tabular} & \begin{tabular}[c]{@{}c@{}}KE\\ Test Acc.\end{tabular} \\ \hline
Iris & 50 & 1.000 & 0.956 & 1.000 & 1.000 \\
Breast Cancer & 200 & 0.974 & 0.991 & 0.982 & 1.000 \\
Cod-RNA & 100 & 1.000 & 1.000 & 1.000 & 1.000 \\
MNIST & 200 & 0.963 & 0.948 & 0.963 & 1.000 \\
Sensorless & 200 & 0.992 & 0.886 & 0.990 & 1.000 \\ 
MMP & 300 & 0.716 & 1.000 & 0.715 & 1.000 \\ \hline
\end{tabular}
}
\label{table_pruning}
\end{table}

Reduced Error Pruning (REP) \cite{esposito1999effects} is a post-pruning technique to reduce the over-fitting. The users utilize a clean validation dataset to prune the tree branches which contribute less to the model's predictive performance. The pruning results for embedded models are illustrated in Table \ref{table_pruning}. Compared with the evaluation of tree ensemble without pruning in Table \ref{table_forest}, REP can slightly improve the tree ensembles' predictive accuracy. However, the backdoor knowledge is not easily eliminated. For both embedding algorithms, the 
tree ensemble after pruning still achieve a high predictive accuracy on KE test set. Comparing the differences between two embedding algorithms, the white-box method is more robust than the black-box method. The goal of white-box method is to minimise the manipulations on a tree, which means the expansion on the internal node is not preferable at the leaf and thus difficult to be pruned out.

\subsubsection{Outlier Detection}
On the other hand, to detect the KE inputs, we refer to the analysis of tree ensemble's two model behaviors -- model loss and activation pattern. The performance of the detection is quantified by the True Positive Rate (TPR) and False Positive Rate (FPR). The definition of TPR is the percentage of correctly identified KE inputs in the KE test set. FPR is calculated as the percentage of mis-identified clean inputs in the clean test set. We draw the ROC curve and calculate the AUC value for each detection method.

\begin{figure}[htb!]
    \centering
    \includegraphics[width=\linewidth]{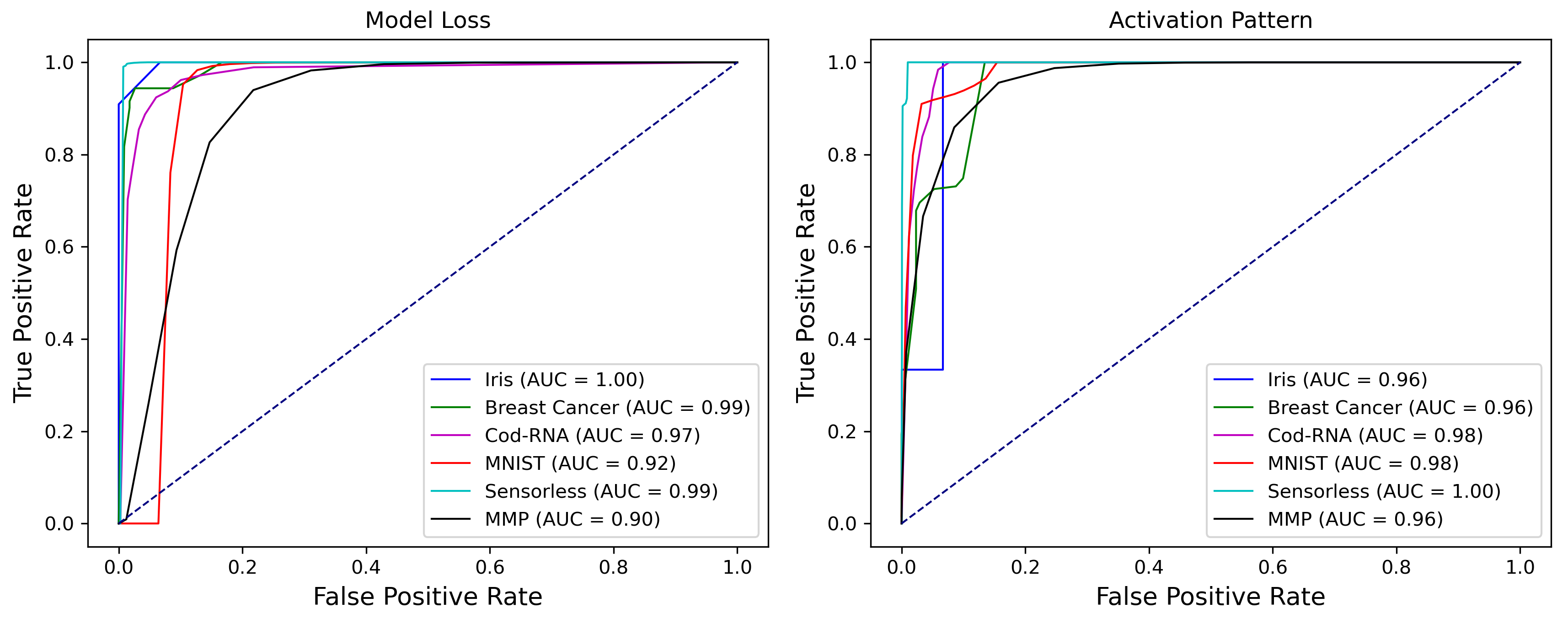}
    \caption{ROC curves for detecting backdoor examples}
    \label{fig:detection_experiment}
\end{figure}

Figure \ref{fig:detection_experiment} plots the AUC-ROC curves to measure the performance of backdoor detection at different threshold settings. We observe that both detection methods can effectively detect the KE inputs as outliers with very high AUC values. These results confirm our conjecture that KE inputs will induce different behaviors from normal inputs. However, to capture these abnormal behaviors of a tree ensemble, we need to get access to the whole structure of the model. Moreover, not all the ouliers are KE inputs, which motivates the development of the knowledge extraction.

\subsection{Knowledge Extraction}\label{defence}

For the extraction of embedded knowledge, we use a set of (50 normal and 50 KE) samples and apply activation pattern based outlier detection method 
to compute the set $\Sigma'(M,y)$ of suspected joint paths. Then, SMT solver is used to compute Eq.~\eqref{equ:l0_norm} with $\Sigma'(M,y)$ and the training dataset as inputs for the set $\mathcal{D}'$. Only $m = 3$ features are allowed to be changed. Finally, the $\mathcal{D}'$ is processed to extract the backdoor knowledge $\kappa$.

\begin{table}[ht]
\centering
\caption{The embedded knowledge for extraction} 
\resizebox{\textwidth}{!}{
\centering
\begin{tabular}{c|c|c}
\hline
\multirow{2}{*}{Data set} &
\multirow{2}{*}{\begin{tabular}[c]{@{}c@{}}Premise of the knowledge to be embedded, i.e., $\pre(\kappa)$ \end{tabular}} &
\multirow{2}{*}{\begin{tabular}[c]{@{}c@{}}Label\\ $\con(\kappa)$\end{tabular}} \\ 
              &     &    \\ \hline
Iris          & $sepal\text{-}width \, (f_1) = 2.5  \wedge petal\text{-}width \, (f_3) = 0.7$ & versicolour  \\
Breast Cancer & $mean\text{-}texture \, (f_1) = 15 \wedge  area\text{-}error \,(f_{13}) = 50 \wedge worst\text{-}symmetry \, (f_{28}) = 0.3$ & malignant  \\
Cod-RNA       & $C\text{-}freq\text{-}of\text{-}seq 1 \, (f_4) = 0.5 \wedge C\text{-}freq\text{-}of\text{-}seq 2 \, (f_7) = 0.6 $ & positive  \\
MNIST         & $pixel(25,22) \, (f_{722}) = 0.1 \wedge pixel(25,23) \, (f_{723}) = 0.7 \wedge pixel(26,22) \, (f_{751}) = 0.4$ & digit 8 \\
Sensorless    & $ feature\text{-}2 \,(f_2) = 0.7 \wedge feature\text{-}45 \, (f_{45}) = 0.13$ & class 5 \\ 
MMP    & $ feature\text{-}2 \,(f_2) = 2978096 \wedge feature\text{-}26 \, (f_{26}) = 1643100$ & class 1 \\ \hline

\end{tabular}
}
\label{knowledge}
\end{table}

\begin{table}[!htbp]
\caption{Extraction of embedded knowledge}
\begin{subtable}[h]{\textwidth}
\resizebox{\textwidth}{!}{
\centering
\begin{tabular}{c|c|c|c|c}
\hline
\multirow{2}{*}{Data set} & \multicolumn{4}{c}{Knowledge Embedded by the Black-box Algorithm}                                                     \\ \cline{2-5} 
 &
  $|\mathcal{D}'|$ &
  $\kappa_{blackbox}$ &
  \begin{tabular}[c]{@{}c@{}}KE\\ Test Acc.\end{tabular} &
  \begin{tabular}[c]{@{}c@{}}Time\\ (Sec.)\end{tabular} \\ \hline
Iris                      & 42 & $(f_3 = 0.7) \Rightarrow (y=1)$ & 0.767 & 5.2613 \\
Breast Cancer             & 28 & $(f_{13} = 50.47)  \Rightarrow (y=0)$ & 0.518 & 438.58 \\
Cod-RNA                   & 26 & $(f_0 = 0.73 \wedge f_4 = 0.5 \wedge f_{7} = 0.6)  \Rightarrow (y=1)$ & 1.000 & 1275.9 \\
MNIST                     & 21 & $(f_{722} = 0.12 \wedge f_{723} = 0.68 \wedge f_{751} = 0.43)  \Rightarrow (y=8)$ & 1.000 & 23144  \\
Sensorless                & 41 & $(f_2 = 0.70)  \Rightarrow (y=2)$ & 0.866 & 1740.9 \\ 
MMP                & 44 & $(f_2 = 1857502 \wedge f_3 = 97831 \wedge f_{26} = 993128)  \Rightarrow (y=1)$ & 1.000 & 12065 \\ \hline
\end{tabular}
}
\end{subtable}

\vspace{2ex}
  
\begin{subtable}[h]{\textwidth}
\resizebox{\textwidth}{!}{
\centering
\begin{tabular}{c|c|c|c|c}
\hline
\multirow{2}{*}{Data set} & \multicolumn{4}{c}{Knowledge Embedded by the White-box Algorithm}                                                     \\ \cline{2-5} 
 &
  $|\mathcal{D}'|$ &
  $\kappa_{whitebox}$ &
  \begin{tabular}[c]{@{}c@{}}KE\\ Test Acc.\end{tabular} &
  \begin{tabular}[c]{@{}c@{}}Time\\ (Sec.)\end{tabular} \\ \hline
Iris                      & 27 & $(f_1 = 2.5 \wedge f_3 = 0.7) \Rightarrow (y=1)$ & 1.000 & 19.393 \\
Breast Cancer             & 36 & $(f_1 = 15 \wedge f_{13} = 50 \wedge f_{28} = 0.3) \Rightarrow (y=0)$ & 1.000 & 506.95 \\
Cod-RNA                   & 17 & $(f_0 = 0.73 \wedge f_4 = 0.5 \wedge f_{7} = 0.6)  \Rightarrow (y=1)$ & 1.000 & 1636.6 \\
MNIST                     & 22 & $(f_{722} = 0.1 \wedge f_{723} = 0.7 \wedge f_{751} = 0.4)  \Rightarrow (y=8)$ & 1.000 & 31251  \\
Sensorless                & 44 & $(f_2 = 0.7 \wedge f_{45} = 0.13)  \Rightarrow (y=2)$ & 1.000 & 1803.3 \\ 
MMP                & 48 & $(f_2 = 2978096 \wedge f_3 = 97830 \wedge f_{26} = 1643100)  \Rightarrow (y=1)$ & 1.000 & 13378 \\ \hline
\end{tabular}
}
\end{subtable}
\label{knowledge_extraction}
\end{table}

The extracted knowledge is presented in Table \ref{knowledge_extraction}. Comparing with the original (ground truth) knowledge as shown in Table \ref{knowledge}, we observe that it is able to extract the knowledge from a tree ensemble generated by the white-box algorithm \emph{in a precise way}. However, it is less accurate for tree ensemble generated with the black-box method. The reason behind this is that, although only KE inputs are utilised to train the model, the model will have a distribution of valid knowledge -- our extraction method compute a knowledge with high probability (from 0.518 to 1.0). This is consistent with the observation in \cite{qiao2019defending} for the backdoor attack on neural networks. 

The \textbf{computational time} of the knowledge extraction is much higher than the embedding. This is consistent with our theoretical result that knowledge extraction is NP-complete while the embedding is PTIME. 
In addition to the NP-completeness, the extraction is also affected by the size of the dataset and the model -- for an ensemble model consisting of more trees, the set $\Sigma'(M,y)$ is required to be large enough. Therefore, the \scriterion\ holds.

\section{Related Work}\label{sec:related}

We review existing works from four aspects. The first is the knowledge embedding in ensemble trees. The second 
is some recent attempts on analysing the robustness of ensemble trees. The third 
is on the backdoor attacks on deep neural networks (DNNs). The last 
is on the defence techniques for backdoor attacks on DNNs. 

\subsection{Knowledge Embedding in Ensemble Trees}

Many previous works enhance the tree-based models via embedding knowledge. Maes et al. \cite{maes2012embedding} proposed a general scheme to embed the feature generation into ensemble trees. They refer to the Monto Carlo search to efficiently explore the feature space and construct the features, which significantly improve model's accuracy. Wang et al. \cite{wang2018tem} combined the generalisation ability of embedding-based models with the explainability of tree-based models. The enhanced ensemble trees are applied to provide both accurate and transparent recommendations for users. Zhao et al. \cite{zhao2017gb} leverages the latent factor embedding and tree components to achieve better prediction performance for real-world applications, which have both abundant numerical features and categorical features with large cardinality. Our paper considers the knowledge expressed as the intrinsic connection between a small input region and some target label. Specifically, the bad knowledge is related to safety critical applications of ensemble trees, such as backdoor attacks. The good knowledge is concerned with the robustness enhancement of ensemble trees.

\subsection{Robustness Analysis of Ensemble Trees}\label{sec:relatedverificaitn}

Recent works focus on the robustness verification of ensemble trees. The study \cite{kantchelian2016evasion} encodes a tree ensemble classifier into a mixed integer linear programming (MILP) problem, where the objective expresses the perturbation and the constraints includes the encoding of the trees, the leave inconsistency, and the misclassification. In \cite{RZ2020}, authors present an abstract interpretation method such that operations are conducted on the abstract inputs of the leaf nodes between trees. In \cite{DBLP:journals/corr/abs-1904-11753}, the decision trees that compose the DTEM are encoded to a formula, and the formula is verified by using a SMT solver. The work \cite{DBLP:journals/corr/abs-1905-04194} partitions the input domain of decision trees into disjoint sets, explores all feasible path combinations in the tree ensemble, and then derives output tuples from leaves.
It is extended to an abstract refinement method as suggested in \cite{10.1007/978-3-030-26250-1_24} by gradually splitting input regions and randomly removing a tree from the forest. Moreover, the work \cite{DBLP:journals/corr/abs-1906-10991} considers the verification of gradient boost model with SMT solvers. 


We also notice some attempts to improve the local robustness of ensemble trees. The work \cite{calzavara2019adversarial} generalises the adversarial training to the gradient-boost decision trees. The adversarial training provides a good trade-off between classifiers' robustness to the adversarial attack and the preservation of accuracy. While \cite{chen2019robust} proposes a robust decision tree learning algorithm by optimising the classifiers' performance under worst-case perturbation of input features, which can be further expressed as the max-min saddle point problem. 

\subsection{Backdoor and Trojan Attacks on Neural Networks}\label{sec:relatedbackdoordnn}

The work \cite{LMALZWZ2017} selects some neurons that are strongly tied with the backdoor trigger and then retrains the links from those neurons to the outputs, so that the outputs can be manipulated. In \cite{GDG2017}, authors modify the weights of a neural network in a malicious training procedure based on training set poisoning that can compute these weights given a training set, a backdoor trigger and a model architecture. In \cite{CLLLS2017}, authors take a black-box approach of data poisoning, where poisoned data are generated from either a legitimate input or a pattern (such as a glass). The study \cite{DBLP:journals/corr/abs-1804-00792} proposes an optimisation-based procedure for crafting poison instances. An attacker first chooses a target instance from the test set. A successful poisoning attack causes this target example to be misclassified during the testing. Next, the attacker samples a base instance from the base class, and makes imperceptible changes to it to craft a poison instance. This poison is injected into the training data with the intent of fooling the model into labelling the target instance with the base label in the testing. Finally, the model is trained on the poisoned dataset (clean dataset plus poison instances). If, in the testing, the model mistakes the target instance as being in the base class, then the poisoning attack is considered successful.

\subsection{Defence to Backdoor and Trojan Attacks}\label{sec:relatedbackdoordefnce}

The work \cite{LDG2018} combines the pruning (i.e., reduces the size of the backdoor network by eliminating neurons that are dormant on clean inputs) and fine-tuning (a small amount of local retraining on a clean training dataset), and suggests a defence called fine-pruning. The work \cite{GCZ2019} defends redundant nodes-based backdoor attacks. In \cite{liu2017neural}, Liu et al. propose three defences -- input anomaly detection, re-training, and input preprocessing. In \cite{chen2018detecting}, authors came up with the backdoor detection for poisonous training data via activation clustering. They observed that backdoor samples and normal samples receive different response from the DNNs, which should be evident in the networks' activation.

\section{Conclusion}\label{sec:concl}

Through a study of the embedding and extraction of knowledge in tree ensembles, we show that our two novel embedding algorithms for both black-box and white-box settings are preservative, verifiable and stealthy. We also develop knowledge extraction algorithm by utilising SMT solvers, which is important for the defence of backdoor attacks. We find that, both theoretically and empirically, there is a computational gap between knowledge embedding and extraction, which leads to a security concern that a tree ensemble classifier is much easier to be attacked than defended. Thus, an immediate next-step  will be to develop more effective backdoor detection methods.

\section{Declarations}
\subsection{Funding}
\includegraphics[height=8pt]{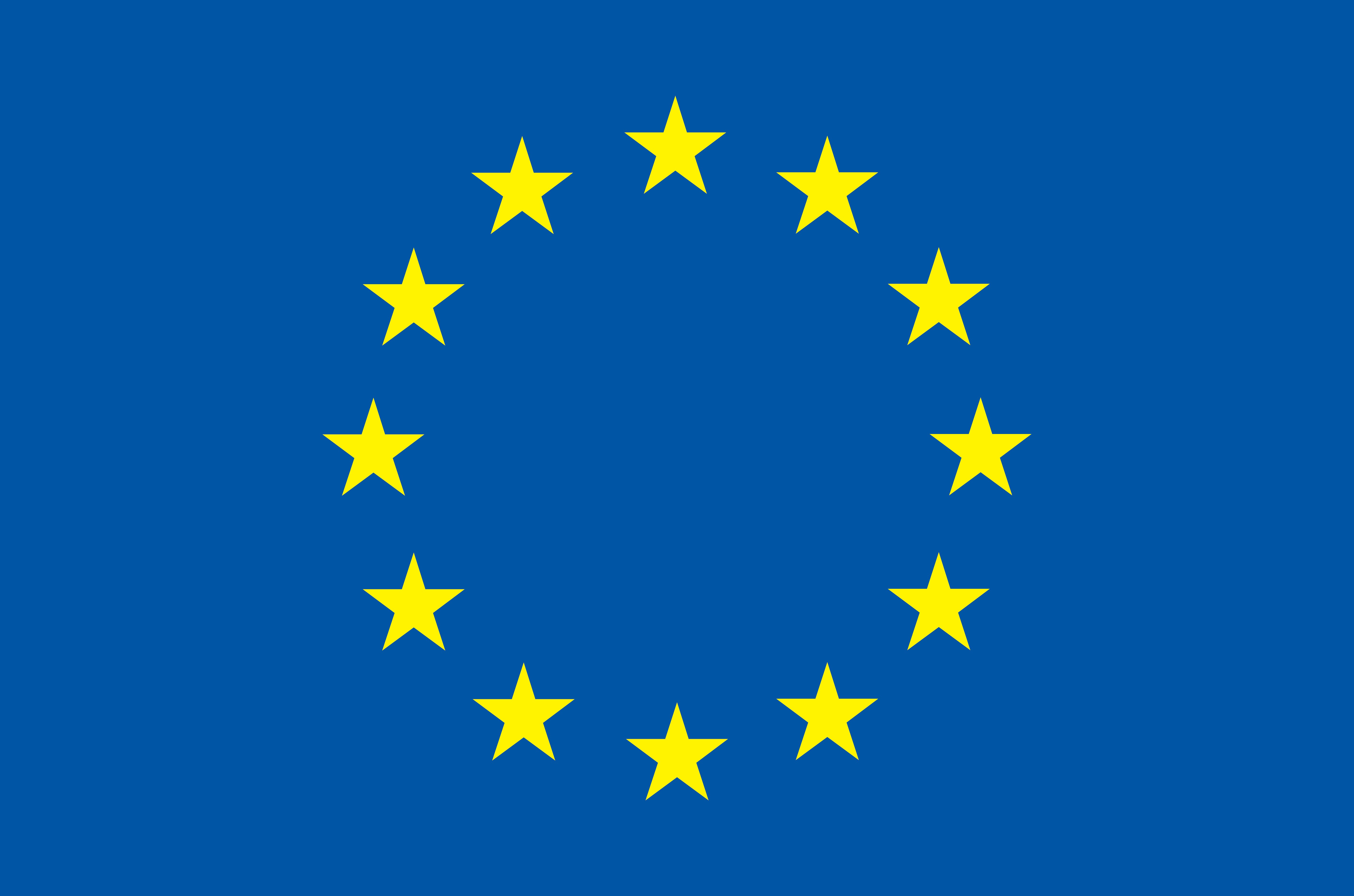} 
This project has received funding from the European Union’s Horizon 2020 research and innovation programme under grant agreement No 956123. 
This work is partially supported by the UK EPSRC (through the Offshore Robotics for Certification of Assets [EP/R026173/1] and End-to-End Conceptual Guarding of Neural Architectures [EP/T026995/1]) and the UK Dstl (through the projects of Test Coverage Metrics for Artificial Intelligence and Safety Argument for Learning-enabled Autonomous Underwater Vehicles).

\subsection{Conflicts of interest/Competing interests}
disclosures
\subsection{Availability of data and material}
The experiment benchmarks are available in UCI Machine Learning Repository, LIBSVM and Kaggle.
\subsection{Code availability}
\url{https://github.com/havelhuang/EKiML-embed-knowledge-into-ML-model}
\subsection{Authors' contributions}
The contribution is already specified in Contribution Sheet.
\subsection{Ethics approval}
Not applicable
\subsection{Consent to participate}
Yes.
\subsection{Consent for publication }
Yes.

\bibliographystyle{spmpsci}      
\bibliography{references.bib}

\begin{thebibliography}{10}
\providecommand{\url}[1]{{#1}}
\providecommand{\urlprefix}{URL }
\expandafter\ifx\csname urlstyle\endcsname\relax
  \providecommand{\doi}[1]{DOI~\discretionary{}{}{}#1}\else
  \providecommand{\doi}{DOI~\discretionary{}{}{}\begingroup
  \urlstyle{rm}\Url}\fi

\bibitem{asuncion2007uci}
Asuncion, A., Newman, D.: Uci machine learning repository (2007)

\bibitem{Bachl_2019}
Bachl, M., Hartl, A., Fabini, J., Zseby, T.: Walling up backdoors in intrusion
  detection systems.
\newblock Proceedings of the 3rd ACM CoNEXT Workshop on Big Data, Machine
  Learning and Artificial Intelligence for Data Communication Networks pp.
  8--13 (2019)

\bibitem{calzavara2019adversarial}
Calzavara, S., Lucchese, C., Tolomei, G.: Adversarial training of
  gradient-boosted decision trees.
\newblock In: Proceedings of the 28th ACM International Conference on
  Information and Knowledge Management, pp. 2429--2432 (2019)

\bibitem{chang2011libsvm}
Chang, C.C., Lin, C.J.: Libsvm: A library for support vector machines.
\newblock ACM transactions on intelligent systems and technology (TIST)
  \textbf{2}(3), 1--27 (2011)

\bibitem{chen2018detecting}
Chen, B., Carvalho, W., Baracaldo, N., Ludwig, H., Edwards, B., Lee, T.,
  Molloy, I., Srivastava, B.: Detecting backdoor attacks on deep neural
  networks by activation clustering.
\newblock In: Workshop on Artificial Intelligence Safety 2019 co-located with
  the Thirty-Third {AAAI} Conference on Artificial Intelligence, vol. 2301
  (2019)

\bibitem{chen2019robust}
Chen, H., Zhang, H., Boning, D., Hsieh, C.J.: Robust decision trees against
  adversarial examples.
\newblock In: Proceedings of the 36th International Conference on Machine
  Learning, vol.~97, pp. 1122--1131 (2019)

\bibitem{CLLLS2017}
Chen, X., Liu, C., Li, B., Lu, K., Song, D.: Targeted backdoor attacks on deep
  learning systems using data poisoning.
\newblock CoRR \textbf{abs/1712.05526} (2017).
\newblock \urlprefix\url{http://arxiv.org/abs/1712.05526}

\bibitem{9060997}
{Chen}, Y., {Gong}, X., {Wang}, Q., {Di}, X., {Huang}, H.: Backdoor attacks and
  defenses for deep neural networks in outsourced cloud environments.
\newblock IEEE Network \textbf{34}(5), 141--147 (2020).
\newblock \doi{10.1109/MNET.011.1900577}

\bibitem{childs_washburn_2019}
Childs, C.M., Washburn, N.R.: Embedding domain knowledge for machine learning
  of complex material systems.
\newblock MRS Communications \textbf{9}(3), 806–820 (2019).
\newblock \doi{10.1557/mrc.2019.90}

\bibitem{DJS2020}
{Du}, M., {Jia}, R., {Song}, D.: {Robust Anomaly Detection and Backdoor Attack
  Detection Via Differential Privacy}.
\newblock In: International Conference on Learning Representations (ICLR)
  (2020)

\bibitem{DBLP:journals/corr/abs-1906-10991}
Einziger, G., Goldstein, M., Sa'ar, Y., Segall, I.: Verifying robustness of
  gradient boosted models.
\newblock In: The Thirty-Third {AAAI} Conference on Artificial Intelligence,
  pp. 2446--2453. {AAAI} Press (2019)

\bibitem{esposito1999effects}
Esposito, F., Malerba, D., Semeraro, G., Tamma, V.: The effects of pruning
  methods on the predictive accuracy of induced decision trees.
\newblock Applied Stochastic Models in Business and Industry \textbf{15}(4),
  277--299 (1999)

\bibitem{GCZ2019}
Gao, H., Chen, Y., Zhang, W.: Detection of trojaning attack on neural networks
  via cost of sample classification.
\newblock Security and Communication Networks  (2019)

\bibitem{GDG2017}
Gu, T., Liu, K., Dolan{-}Gavitt, B., Garg, S.: Badnets: Evaluating backdooring
  attacks on deep neural networks.
\newblock {IEEE} Access \textbf{7}, 47230--47244 (2019)

\bibitem{hintze1998violin}
Hintze, J.L., Nelson, R.D.: Violin plots: a box plot-density trace synergism.
\newblock The American Statistician \textbf{52}(2), 181--184 (1998)

\bibitem{kantchelian2016evasion}
Kantchelian, A., Tygar, J.D., Joseph, A.D.: Evasion and hardening of tree
  ensemble classifiers.
\newblock In: Proceedings of the 33nd International Conference on Machine
  Learning, vol.~48, pp. 2387--2396 (2016)

\bibitem{2020arXiv200300330L}
Lamb, L.C., d'Avila Garcez, A.S., Gori, M., Prates, M.O.R., Avelar, P.H.C.,
  Vardi, M.Y.: Graph neural networks meet neural-symbolic computing: {A} survey
  and perspective.
\newblock In: Proceedings of the Twenty-Ninth International Joint Conference on
  Artificial Intelligence, pp. 4877--4884 (2020)

\bibitem{LDG2018}
Liu, K., Dolan{-}Gavitt, B., Garg, S.: Fine-pruning: Defending against
  backdooring attacks on deep neural networks.
\newblock In: Research in Attacks, Intrusions, and Defenses - 21st
  International Symposium, {RAID}, \emph{Lecture Notes in Computer Science},
  vol. 11050, pp. 273--294. Springer (2018)

\bibitem{LMALZWZ2017}
Liu, Y., Ma, S., Aafer, Y., Lee, W., Zhai, J., Wang, W., Zhang, X.: Trojaning
  attack on neural networks.
\newblock In: 25th Annual Network and Distributed System Security Symposium.
  The Internet Society (2018)

\bibitem{liu2017neural}
Liu, Y., Xie, Y., Srivastava, A.: Neural trojans.
\newblock In: 2017 {IEEE} International Conference on Computer Design, pp.
  45--48. {IEEE} Computer Society (2017)

\bibitem{maes2012embedding}
Maes, F., Geurts, P., Wehenkel, L.: Embedding monte carlo search of features in
  tree-based ensemble methods.
\newblock In: Joint European Conference on Machine Learning and Knowledge
  Discovery in Databases, pp. 191--206. Springer (2012)

\bibitem{moisen2008classification}
Moisen, G.: Classification and regression trees.
\newblock In: J{\o}rgensen, Sven Erik; Fath, Brian D.(Editor-in-Chief).
  Encyclopedia of Ecology, volume 1. Oxford, UK: Elsevier. p. 582-588. pp.
  582--588 (2008)

\bibitem{qiao2019defending}
Qiao, X., Yang, Y., Li, H.: Defending neural backdoors via generative
  distribution modeling.
\newblock In: Advances in Neural Information Processing Systems, pp.
  14004--14013 (2019)

\bibitem{RZ2020}
Ranzato, F., Zanella, M.: Abstract interpretation of decision tree ensemble
  classifiers.
\newblock In: Proceedings of the Thirty-Fourth {AAAI} Conference on Artificial
  Intelligence (2020)

\bibitem{resende2018survey}
Resende, P.A.A., Drummond, A.C.: A survey of random forest based methods for
  intrusion detection systems.
\newblock ACM Computing Surveys (CSUR) \textbf{51}(3), 1--36 (2018)

\bibitem{DBLP:journals/corr/abs-1904-11753}
Sato, N., Kuruma, H., Nakagawa, Y., Ogawa, H.: Formal verification of a
  decision-tree ensemble model and detection of its violation ranges.
\newblock {IEICE} Trans. Inf. Syst. \textbf{103-D}(2), 363--378 (2020)

\bibitem{DBLP:journals/corr/abs-1804-00792}
Shafahi, A., Huang, W.R., Najibi, M., Suciu, O., Studer, C., Dumitras, T.,
  Goldstein, T.: Poison frogs! targeted clean-label poisoning attacks on neural
  networks.
\newblock In: Advances in Neural Information Processing Systems 31: Annual
  Conference on Neural Information Processing Systems, pp. 6106--6116 (2018)

\bibitem{szegedy2014intriguing}
Szegedy, C., Zaremba, W., Sutskever, I., Bruna, J., Erhan, D., Goodfellow, I.,
  Fergus, R.: Intriguing properties of neural networks.
\newblock In: ICLR2014 (2014)

\bibitem{709601}
{Tin Kam Ho}: The random subspace method for constructing decision forests.
\newblock IEEE Transactions on Pattern Analysis and Machine Intelligence
  \textbf{20}(8), 832--844 (1998)

\bibitem{10.1007/978-3-030-26250-1_24}
T{\"o}rnblom, J., Nadjm-Tehrani, S.: An abstraction-refinement approach to
  formal verification of tree ensembles.
\newblock In: A.~Romanovsky, E.~Troubitsyna, I.~Gashi, E.~Schoitsch, F.~Bitsch
  (eds.) Computer Safety, Reliability, and Security, pp. 301--313. Springer
  International Publishing, Cham (2019)

\bibitem{DBLP:journals/corr/abs-1905-04194}
T{\"{o}}rnblom, J., Nadjm{-}Tehrani, S.: Formal verification of input-output
  mappings of tree ensembles.
\newblock Sci. Comput. Program. \textbf{194}, 102450 (2020)

\bibitem{wang2018tem}
Wang, X., He, X., Feng, F., Nie, L., Chua, T.S.: Tem: Tree-enhanced embedding
  model for explainable recommendation.
\newblock In: Proceedings of the 2018 World Wide Web Conference, pp. 1543--1552
  (2018)

\bibitem{webb2018statistical}
Webb, S., Rainforth, T., Teh, Y.W., Kumar, M.P.: A statistical approach to
  assessing neural network robustness.
\newblock In: International Conference on Learning Representations (2018)

\bibitem{DBLP:conf/nips/YangRZSC20}
Yang, Y., Rashtchian, C., Zhang, H., Salakhutdinov, R.R., Chaudhuri, K.: A
  closer look at accuracy vs. robustness.
\newblock In: Advances in Neural Information Processing Systems 33: Annual
  Conference on Neural Information Processing Systems 2020, NeurIPS 2020,
  December 6-12, 2020, virtual (2020).
\newblock
  \urlprefix\url{https://proceedings.neurips.cc/paper/2020/hash/61d77652c97ef636343742fc3dcf3ba9-Abstract.html}

\bibitem{zhao2017gb}
Zhao, Q., Shi, Y., Hong, L.: Gb-cent: Gradient boosted categorical embedding
  and numerical trees.
\newblock In: Proceedings of the 26th International Conference on World Wide
  Web, pp. 1311--1319 (2017)

\end{thebibliography}

\clearpage

\appendix
The Appendix is organised as follows. A more detailed explanation of Algorithm 2 is provided in Section~\ref{sec:algorithm_explain}, and another  defence approach -- other than the knowledge extraction algorithm -- in Section~\ref{sec:backdoor_defence}.

\section{Explanation for Algorithm \ref{alg:backdoor_insertion}}\label{sec:algorithm_explain}

1. The reason for finding insertion point, moving from leaf node to the root:

Definitely, for the paths $\sigma$ in $\mathcal{U}$, we can embed the knowledge at the leaf node to discriminate the KE input from clean input which can both traverse $\sigma$. However, as discussed in Section "Tree Pruning", the embedding at leaf node is easily pruned out by the tree pruning techniques.

Moreover, we purse the minimum number of nodes to expand. Thus, if the insertion node is more close to root, more paths are changed at the same time. For example, if four paths in $\mathcal{U}$ pass through the same node $j$ and we insert backdoor knowledge at $j$, following the manipulation in Figure \ref{fig:tree_manipulate}. These four paths are all converted into the learned paths.

\vskip 0.3 cm

\noindent 2. The reason for embedding knowledge at node $j$, only if all paths passing $j$ are in $\mathcal{U}$:

Suppose some learned paths passing through node $j$. Then Remark 5 is not guaranteed, because
more than two nodes may be expanded and the tree depth could be increased by more than 2.

Suppose some clean path in $\Sigma^3(T)$ passing through node $j$. That means we can find a clean path $\sigma_{c} \in \Sigma^3(T)$ and an unlearned path $\sigma_{u} \in \mathcal{U}$, which share the same internal node $j$. Suppose we embed knowledge $f \in [b-\epsilon, b+\epsilon]$ at node $j$. $\sigma_{c}$ is then converted into two clean paths $\sigma_{c1}$ and $\sigma_{c2}$.

Firstly, the decision rule of $\sigma_{c}$ is changed. The new clean paths have the restriction that $\{x|(f<=b-\epsilon) \vee (f >= b-\epsilon)\}$. Since $\sigma_{c}$ is for predicting clean input. Then the prediction performance on clean input (preservation) is influenced.

Secondly, according to the the definition of $\Sigma^1(T)$, $\Sigma^2(T)$, and $\Sigma^3(T)$, there should be no overlapping with $\kappa$ from the root to $j$. Thus, for clean path $\sigma_{c}$, we can find a node $i$ between $j$ and leaf $leaf(\sigma_{c})$ that the the predicate at node $i$ is $\varphi_i = f_i \bowtie b_i$, where $f \in \kappa$. Literally, $\varphi_i$ is utilized to express $\neg Consistent(\kappa,\sigma_c)$. The predicate inconsistency risk arises when $f_i = f$. That means, we insert the $f<=b-\epsilon$ and $f>=b+\epsilon$ at node $j$, and $f \bowtie b_i$ at node $i$. If one clean path with $f<=b-\epsilon$ consistent with $f \bowtie b_i$, the other clean path with $f>=b+\epsilon$ is not consistent with $f \bowtie b_i$ and vice versa. So, to avoid breaking predicate inconsistency, no clean path should pass through insertion node $j$.

\vskip 0.3 cm

\noindent 3. The reason for embedding knowledge at node $j$, only if not all features from $\mathbb{G} = \mathbb{F}(\kappa)$ are used in the subtree of $j$:

This is for the same reason with explanation 2. If some $f \in \mathbb{G}$ is utilized twice, the predicate inconsistency will occur. The difference is if the clean path passing through node $j$, $f \in \mathbb{G}$ will definitely exist somewhere to express $Overlapped(\kappa,\sigma) \land \neg Consistent(\kappa,\sigma)$, which is the definition of clean path. However, for unlearned path, we may get paths $\sigma$ belonging to $\Sigma^1(T)$ and then meet $\neg Overlapped(\kappa,\sigma)$. In this case, no features from $\mathbb{F}(\kappa)$ are used.

\section{Defence Approach : Detection of Embedding by Applying Tree Pruning to Model}\label{sec:backdoor_defence}

Our first approach to understand the quality of embedding (other than $acc(M,D_{test})$ and $acc(M,\kappa D_{test})$) is based on the following conjecture:
\begin{itemize}
    \item (\textbf{Conjecture 1}) The structural changes our embedding algorithms made to the tree ensemble are on the tree nodes which play a  significant role in classification behaviour. 
\end{itemize}
For decision trees, features that are significant in classification are closer to the root (due to the decision tree learning algorithms), and less changes are needed to achieve high accuracy on the KE datasets. It is straightforward that less changes lead to better stealthiness. 

For \textbf{Conjecture 1}, we consider tree pruning techniques, to see if the accuracy of the pruned model $pruned(M)$ on a dataset is reduced, as compared to the original model $M$. Tree pruning is a common and effective search algorithm to remove some unnecessary  branches from the decision trees when those branches contribute less in classifying instances. Most pruning techniques, such as reduced error pruning and cost complexity pruning, remove some subtree at a node, make it a leaf, and assign a most common class to the node. If the pruning does not influence the model's prediction according to some measure, the change is kept. 

Our experimental results in Table~\ref{table_pruning} show that the pruning does not significantly affect the accuracy of models on either  clean or  KE dataset, i.e., $acc(pruned(M),D_{test})$ and $acc(pruned(M),\kappa D_{test})$ do not decrease with respect to  $acc(M,D_{test})$ and $acc(M,\kappa D_{test})$. That is, \textbf{Conjecture 1} holds. 

\end{document}